\documentclass[journal]{IEEEtran}
\usepackage[dvips]{graphicx}
\usepackage{amsmath}
\usepackage{amssymb}
\usepackage{epic}
\allowdisplaybreaks

\newcommand{\qed}{\hspace*{\fill} $\Box$ \\}

\newtheorem{theorem}{Theorem}
\newtheorem{corollary}{Corollary}

\newtheorem{lemma}{Lemma}

\begin{document}

\title{Asymptotically Optimal Multiple-access\\ Communication via Distributed Rate Splitting}
%\markboth{Submitted to \textit{IEEE Transactions on Information Theory}}{Cao and Yeh:Distributed Rate
%Splitting in Gaussian and Discrete Memoryless Multiple-access Channels}
\author{Jian~Cao,~\IEEEmembership{Student Member,~IEEE,}
        and~Edmund~M.~Yeh,~\IEEEmembership{Member,~IEEE}% <-this % stops a space
\thanks{Manuscript received April 19, 2005; revised September 21, 2006.
        This work was supported in part by Army Research Office (ARO) Young Investigator Program (YIP) grant DAAD19-03-1-0229 and
         by National Science Foundation (NSF) grant CCR-0313183. The material in this paper was presented in part at the Allerton Conference
on Communication, Control, and Computing, Monticello, IL, September, 2004, the IEEE International Symposium on
Information Theory, Adelaide, Australia, September, 2005, and the Global Telecommunications Conference, St. Louis,
MO, November, 2005.}% <-this % stops a space
\thanks{The authors are with the Department of Electrical Engineering, Yale University, New Haven, CT 06520, U.S.A. (Email:jian.cao@yale.edu; edmund.yeh@yale.edu).}%
\thanks{Communicated by Y. Steinberg, Associate Editor for Shannon Theory.}}

\maketitle

\begin{abstract}
We consider the multiple-access communication problem in a distributed setting for both the additive white Gaussian
noise channel and the discrete memoryless channel.  We propose a scheme called Distributed Rate Splitting to achieve
the optimal rates allowed by information theory in a distributed manner.  In this scheme, each real user creates a
number of virtual users via a power/rate splitting mechanism in the $M$-user Gaussian channel or via a random switching
mechanism in the $M$-user discrete memoryless channel. At the receiver, all virtual users are successively decoded.
Compared with other multiple-access techniques, Distributed Rate Splitting can be implemented with lower complexity and
less coordination.  Furthermore, in a symmetric setting, we show that the rate tuple achieved by this scheme converges
to the maximum equal rate point allowed by the information-theoretic bound as the number of virtual users per real user
tends to infinity.  When the capacity regions are asymmetric, we show that a point on the dominant face can be achieved
asymptotically.  Finally, when there is an unequal number of virtual users per real user, we show that differential
user rate requirements can be accommodated in a distributed fashion.
\end{abstract}


\section{Introduction} \label{sec:intro}
We consider the basic multiple-access communication problem in a
distributed setting. In Gallager's survey paper
\cite{paper:gal85}, it is pointed out that the multiple-access
 problem has been studied from a number of different
perspectives, each having its own advantages and shortcomings. In the data networking community, a well-known
distributed multiple access scheme is ALOHA~\cite{paper:abr70,book:bg92}. In ALOHA, it is assumed that a ``collision"
happens whenever more than one user transmit simultaneously. Those packets involved in a collision are discarded and
retransmitted according to some retransmission probability. The collision channel model, however, does not accurately
describe the underlying physical multiple-access channel. It is well known that there exist coding techniques which can
decode multiple users' messages when simultaneous transmissions occur. Indeed, more sophisticated models such as signal
capture \cite{paper:abr77,paper:nam84}, spread ALOHA \cite{paper:mm92}, and multi-packet reception
\cite{paper:gvs89,paper:nmt03,paper:ray81} have been developed to enhance ALOHA. Even these improved schemes, however,
are not optimal from the viewpoint of information theory. In the information theory literature, the capacity regions of
various multiple-access channel (MAC) models have been characterized (see \cite{paper:wyn74,book:ct91,paper:ahl71,
thesis:lia72, paper:pol83, paper:hh85}). Rate splitting multiple-access techniques (or generalized time-sharing) are
presented in \cite{paper:ru96, paper:yg98, paper:gruw01, paper:rim01} to achieve every point in the Gaussian or the
discrete memoryless MAC capacity region using only single-user codes. These schemes, however, require a pre-defined
decoding order, which makes distributed implementation difficult. Finally, in the spread spectrum community, CDMA
techniques are adopted. Here, users are decoded regarding all other users' signals as interference. This, however, is
not optimal from the information theoretic viewpoint.

To address some of the shortcomings mentioned above,  Medard
\textit{et al.}~\cite{paper:med04}  use information-theoretic
techniques to analyze different notions of capacity for
time-slotted ALOHA systems. A coding/decoding scheme which
combines rate splitting and superposition coding is constructed.
This scheme allows some bits to be reliably received even when
collision occurs, and more bits to be reliably received in the
absence of collisions. Shamai \cite{paper:sha00} proposes a
similar scheme to apply a broadcast strategy to multiple-access
channel under static fading where the fading coefficients are not
available to the transmitters or the receiver. To implement the
scheme in~\cite{paper:med04}, however, a pre-defined decoding
order is required, as in~\cite{paper:ru96, paper:yg98,
paper:gruw01, paper:rim01}. In~\cite{paper:che96}, Cheng proposes
a distributed scheme called ``stripping CDMA'' for the $L$ out of
$K$ Gaussian MAC. Here, no pre-defined decoding order is required.
It is shown in~\cite{paper:che96} that stripping CDMA is
asymptotically optimal, although the optimal operating parameters
are not specified.

In this paper, we investigate distributed multiple-access schemes based on the idea of rate splitting for both the
$M$-user additive white Gaussian noise MAC and the $M$-user discrete memoryless MAC.  We characterize the optimal
operating parameters as well as the asymptotic optimality of these schemes from the viewpoint of information theory.
Assume that every user has an infinite backlog of bits to send, and that every user knows the total number of users
$M$.  We propose a distributed scheme, called \textit{Distributed Rate Splitting} (DRS), to achieve the optimal
communication rates allowed by information theory. In this scheme, each real user creates a number of virtual users via
a power/rate splitting mechanism in the $M$-user Gaussian channel or via a random switching mechanism in the $M$-user
discrete memoryless channel. At the receiver, all virtual users are successively decoded. A possible advantage of the
DRS scheme is that it can be implemented with lower complexity when compared with multiple-access schemes such as joint
coding\footnote{The lower complexity comes from the fact that the DRS scheme uses single-user codes instead of
multi-user codes. As we show later, the DRS scheme with a reasonable number of virtual users per real user allows us to
get close to the optimal operating rates.} and less coordination when compared with time-sharing and rate splitting.

In Sections \ref{sec:gau} and \ref{sec:dmc}, we focus first on
symmetric situations where the channel capacity regions are
symmetric and every real user creates the same number of virtual
users. In this case, the DRS scheme entails the following. Each
user $i$ creates $L$ virtual users indexed by $i_k$, $k =
1,2,...,L$. The virtual user class $V_k$ consists of users
$\{1_k,...,M_k \}$ (i.e. we have altogether $L$ virtual user
classes and there are $M$ virtual users in each class). In the
$M$-user Gaussian MAC, virtual users are created via a power/rate
splitting mechanism. The signal transmitted by a real user is the
superposition of all its virtual users' signals. The receiver
receives the sum of the virtual users' signals plus noise. All
virtual users are then successively decoded in increasing order of
their class. That is, all virtual users in class $V_k$,
$k=1,...,L$, are decoded before any virtual user in $V_j$, where
$j > k$, is decoded.  In contrast to~\cite{paper:che96}, the
optimal operating parameters, such as power and rate, are
explicitly specified for any finite $L$.  In the $M$-user discrete
memoryless MAC, virtual users with the same input distribution as
the real users are created, and the transmitted signal of a real
user is determined by a random switch. The receiver successively
decodes all virtual users in increasing order of their class given
the side information of already decoded virtual users. The optimal
switch is found for any finite $L$ for the $2$-user case. Finally,
it is shown that for both channel models, the rate tuple achieved
by the DRS scheme converges to the maximum equal rate point
allowed by the information-theoretic bound as the number of
virtual users per real user tends to infinity.

Next, in Section~\ref{sec:variation}, we consider more general
situations where the capacity regions can be asymmetric and real
users may generate different numbers of virtual users. For the
case of asymmetric capacity regions, new operating parameters are
specified  for any finite number of virtual users per real user.
We show that the DRS scheme still can achieve a point on the
dominant face as the number of virtual users per real user tends
to infinity. For the case of unequal number of virtual users per
real user, we present a variation of DRS which supports
differential user rate requirements in a distributed manner. In
this new scheme, each user~$i$, independently from other users,
generates $L_i$ virtual users according to its own rate
requirement. All virtual users are then decoded reliably at the
receiver. Furthermore, as each real user generates more virtual
users, the rate tuple achieved under this variation of DRS
converges the maximum equal rate point on the dominant face.

\section{$M$-user Gaussian Multiple-access Channel} \label{sec:gau}
%\subsection{Optimal Transmission Rate Distribution Function}\label{sec:gauopt}

We first examine a Gaussian MAC with a symmetric capacity region. Later in Section~\ref{sec:variation}, we consider the
asymmetric case. Consider an $M$-user Gaussian MAC where each transmitter has transmission power $P$ and the receiver
has noise variance $N$. The capacity region ${\cal C}$ is the set of $\mathbf{R} = \left(R_1,...,R_M\right) \in
\mathbb{R}_+^M$ satisfying\footnote{We use $\mathbf{R}$ to denote $\left(R_1,...,R_M\right)$ throughout.}
\begin{equation} \label{eq:capacityregion}
\sum_{i \in S} R_i \leq \frac{1}{2} \log \left(1+ \frac{|S|P}{N} \right) \text{ } \forall S \subseteq \left\{1,...,M
\right\},
\end{equation}
where $|S|$ is the cardinality of the set $S$. The \textit{dominant face} $\cal D$ is the subset of rate tuples which
gives equality in \eqref{eq:capacityregion} for $S=\left\{1,...,M\right\}$. For this symmetric setting, it is easy to
see that the maximum common rate that every user can achieve\footnote{$\mathbf{R}$ is achievable if for any
$\varepsilon > 0$, there exists an $\left(n, R_1 - \varepsilon, ... , R_M - \varepsilon \right)$ multiple-access code
with overall error probability $P_e < \varepsilon$, where $n$ is the block length.} is $R^* = \frac{1}{2M} \log
(1+\frac{MP}{N})$.

\begin{figure}
\begin{center}
\includegraphics[width = 6cm]{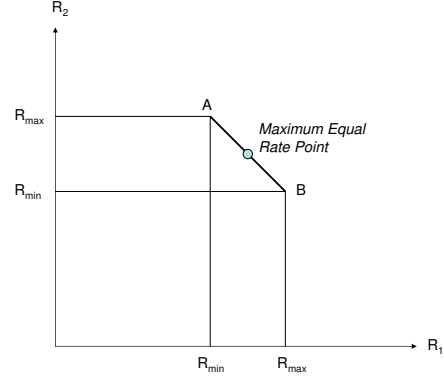} \caption{Two-user Gaussian
multiple-access capacity region, where the dominant face has been highlighted.} \label{fig:twogaussian}
\end{center}
\end{figure}

It is well-known that rate tuples on the dominant face other than
the vertices cannot be achieved via standard successive
decoding~\cite{book:ct91}. Note that the optimal rate tuple
$\mathbf{R}^* \equiv \left(R^*,...,R^*\right)$, called the
\textit{maximum equal rate point}, is such a point. For the
two-user Gaussian MAC, the maximum equal rate point is shown in
Fig.~\ref{fig:twogaussian}.
Currently, three methods are known to achieve general points on the dominant face: joint encoding/decoding,
time-sharing, and rate-splitting. Joint encoding/decoding is not practical because of its high
complexity~\cite{paper:gal85}. In time-sharing, all $M$ users need to coordinate their transmissions. Therefore, some
communication overhead is required. The rate-splitting method in~\cite{paper:ru96} achieves every point in $\cal C$ via
a generalized successive decoding scheme. For the two-user case, user~$1$ creates two virtual users, say $1a$ and $1b$,
by splitting its power $P$ into $\delta$ and $P-\delta$ and setting $r_{1a}= \frac{1}{2} \log (1 +
\frac{\delta}{2P-\delta+N} )$, $r_{1b}= \frac{1}{2} \log (1 + \frac{P- \delta}{N} )$. User $2$ does not split its power
and sets its rate to $R_2 = \frac{1}{2} \log (1 + \frac{P}{P-\delta+N})$. The decoding order is $(1a,2,1b)$. In order
to achieve the maximum equal rate point, we solve $R_2 = r_{1a}+r_{1b}$, yielding $\delta=\frac{1}{2} (N+2P- \sqrt{N
(N+2P)} )$. Thus, both time-sharing and rate splitting require some coordination among users.

In this paper, we focus on {\em distributed} multiple-access communication schemes.  In particular, we introduce the
\textit{Distributed Rate Splitting} (DRS) scheme.  The DRS scheme offers the possibility of multiple-access
communication with lower complexity when compared with joint coding, and communication with less coordination when
compared with the time-sharing or rate splitting method. Moreover, we show that DRS can achieve the maximum equal rate
point of the MAC capacity region asymptotically.

We now formally present the DRS scheme. In this scheme, each user
creates $L$ virtual users by splitting its power $P$ into
$\left(p_1,p_2,...,p_L\right)$, where $p_k$ is the power allocated
to the $k$th virtual user and $\sum_{k=1}^L p_k = P$. Each user
then assigns transmission rate $r_k$ to virtual user $k$. Note
that the proposed DRS scheme is symmetric, i.e. all $M$ users
split their powers and set their rates in the same way. The signal
transmitted by a user is the superposition of its virtual users'
signals. As defined in Section \ref{sec:intro}, virtual user class
$V_k$ consists of all virtual users indexed by $k$. The receiver
receives the sum of all virtual users' signals plus noise. All
virtual users are then successively decoded in increasing order of
their class.

\begin{figure}
\begin{center}
\includegraphics[width = 8cm]{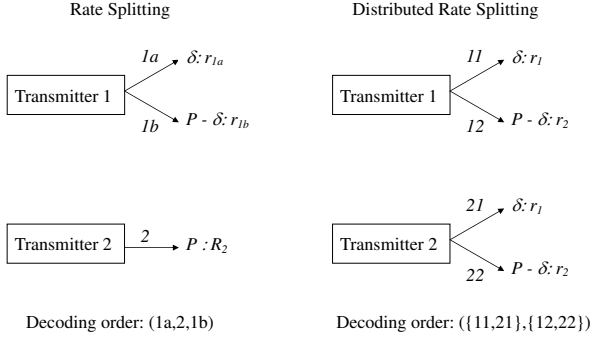}
\caption{Comparison of rate splitting with distributed rate splitting. In the right-hand figure, the virtual user class
$V_1 = \{11,21\}$ is decoded before virtual user class $V_2 = \{12,22\}$.} \label{fig:difference}
\end{center}
\end{figure}

To illustrate the DRS scheme, consider the case $L=2$. Each real
user splits its power $P$ into $\delta$ and $P-\delta$. Notice
there are two major differences between our scheme and the
traditional rate splitting scheme in \cite{paper:ru96}. First, in
our scheme, all real users split in the same way, whereas there is
at least one user who does not split in the traditional rate
splitting scheme. Second, virtual users in the same class, (i.e.
with the same index $k$), are allocated the same rate in our
scheme, whereas all virtual users have different rates according
to the pre-defined decoding order in the traditional rate
splitting scheme.\footnote{In terms of achievable rate, the DRS
scheme with $L=2$ is not optimal. Later in this section, we
demonstrate the asymptotic optimality of DRS by taking $L$ to
infinity.} These differences are illustrated in
Fig.~\ref{fig:difference}.

Since we assume the receiver uses successive decoding method, some virtual user must be decoded first. Without loss of
generality we assume one of the $\delta$ virtual users is decoded first. For the case $L=2$, we show that there is a
unique way for a real user to split its power in order to maximize its total throughput. \vspace{0.2cm}
\begin{lemma}
For $L = 2$ and for a fixed $\delta$, each real user's throughput is maximized by setting $r_{1} = \frac{1}{2}\log ( 1
+ \frac{\delta}{MP-\delta+N} ) $ and $r_{2} = \frac{1}{2} \log ( 1 + \frac{P-\delta}{ ( M-1 ) (P-\delta ) +N} )$.
\end{lemma}

\vspace{0.2cm} \textit{Proof:} The $\delta$ virtual user who is decoded first must have $r_{1} = \frac{1}{2} \log ( 1 +
\frac{\delta}{MP-\delta+N} ) $ (i.e. the virtual user regards all other virtual users as interference) in order to be
decoded successfully. Due to symmetry, all other $\delta$ virtual users must have the same $r_{1}$. Then the problem of
maximizing each real user's throughput reduces to $\max r_{2}$, subject to~(i) $r_{1} = \frac{1}{2} \log ( 1
+\frac{\delta}{MP-\delta+N})$, (ii) one of the $\delta$ virtual users is decoded first and (iii) $
\left(r_{1},r_{2},r_1,r_2,...,r_{1},r_{2}\right) \text{must be decodable.}$

Note that $r_{2}$ is maximized when the interference plus noise faced by all the $\left(P-\delta\right)$ virtual users
is minimized, and the only way to minimize the interference plus noise faced by all the $\left(P-\delta\right)$ virtual
users is to decode all the $\delta$ virtual users before decoding any $\left(P-\delta\right)$ virtual user.\footnote{We
assume a genie-aided \cite{paper:ru96} decoding scheme where the previously decoded messages have been decoded
correctly. In practice, errors can be made in previous decodings. However, for purposes of analyzing the overall error
probability, the genie-aided model is sufficient.} Therefore, the minimum interference plus noise faced by any
$\left(P-\delta\right)$ virtual user is $MP-M\delta - \left(P-\delta\right)+N = \left( M - 1 \right) \left(P-\delta
\right)+N$. Hence, the maximum rate associated with a $\left(P-\delta\right)$ virtual user is $r_{2} = \frac{1}{2} \log
( 1 + \frac{P-\delta}{( M-1 ) (p-\delta ) +N} )$. \qed

Using the DRS scheme with $L = 2$, each user can strictly increase its throughput relative to the case where users do
not split their powers and decode against each other as noise. This is easily verified by observing that for any
$\delta < P$, \begin{multline} \log \left( 1 + \frac{\delta}{MP-\delta+N} \right) + \log \left( 1 +
\frac{P-\delta}{\left(M-1\right) \left(P-\delta \right) + N} \right) \\
  >  \log \left(1 +
\frac{P}{\left(M-1\right)P+N}\right). \label{eq:lemma2}
\end{multline}

Now consider the case where each user creates more than two virtual users $(L > 2)$. Here, we show that each user's
throughput increases further. \vspace{0.2cm}
\begin{lemma} \label{le:morevuser}
Given a DRS scheme with $L$ virtual users per real user, where $\left(p_1,...,p_L\right)$ are the virtual users'
powers, it is possible to strictly increase the throughput via an $(L+1)$ virtual user system with powers
$(p_1,...,p_{L-1},p_L^{'},p_L^{''})$, where $p_L^{'}+p_L^{''}=p_L$.
\end{lemma}

\vspace{0.2cm} \textit{Proof:} Suppose that every user splits its power into $L$ virtual users: $\left(
p_1,p_2,...,p_{L-1},p_L \right)$ subject to $\sum_{k=1}^L p_k = P$, where $L$ is an arbitrary integer and $p_k$ is the
power of $k$th virtual user. Since virtual user $L$ is decoded last, following the reasoning in the proof of Lemma~$1$,
we have $r_L = \frac{1}{2} \log (1+\frac{p_L}{(M-1)p_L+N} )$. We now split the virtual user with power $p_L$ into two
new virtual users with powers $p^{'}_{L}$ and $p^{''}_{L}$, where $p^{'}_{L}+p^{''}_{L} = p_L$. We set
\begin{eqnarray*}
r_{L^{'}} &=& \frac{1}{2} \log \left( 1 + \frac{p^{'}_{L}}{Mp_L-p^{'}_{L}+N} \right)\\
\quad r_{L^{''}} &=& \frac{1}{2} \log \left( 1 + \frac{p^{''}_{L}}{\left(M-1\right)p^{''}_{L}+N} \right).
\end{eqnarray*}

Now each real user has $L+1$ virtual users. Notice that we do not change the power and decoding order of any of the
other virtual users (i.e. virtual users $1,\ldots,L-1$). From a real user's view point, the virtual user with
$r_{L^{'}}$ is decoded second to last \textit{among all virtual users generated by this real user} and the virtual user
with $r_{L^{''}}$ is decoded last. Thus, all virtual users can be decoded and from \eqref{eq:lemma2}, $r_{L^{'}} +
r_{L^{''}}
> r_L$. Therefore, every real user with $L$ virtual users can strictly increase its throughput by splitting its power among $L+1$
virtual users. \qed

Before we examine the asymptotic behavior of DRS, we solve the problem of how to split a user's power
\textit{optimally} among a fixed number of virtual users. The main difficulty here is that the objective function is
not concave. In order to find the optimal splitting method, we prove the following lemma.

\vspace{0.5cm}
\begin{lemma} \label{lemma:maximization}
Consider the following optimization problem:

\begin{equation} \label{eq:optimization}
\max_{p_k,p_j} \frac{1}{2} \log \left( 1 + \frac{p_k}{A-p_k} \right) + \frac{1}{2} \log \left( 1 +
\frac{p_j}{A-Mp_k-p_j}\right) \end{equation} subject to $p_k$+$p_j$ = $c$ and $p_k,p_j \geq 0$, where $A$, $M$ and $c$
are positive constants and $A \geq Mc$. The unique solution to \eqref{eq:optimization} is also the unique solution to
$\frac{p_k}{A-p_k} = \frac{p_j}{A-Mp_k-p_j}, \text{ subject to } p_k+p_j=c \text{ and } p_k,p_j \geq 0$, where $A \geq
Mc$.
\end{lemma}\vspace{0.2cm}

\textit{Proof:} Substitute $p_j = c-p_k$ into the objective function, we have
\begin{eqnarray*}
f\left(p_k\right) &=& \frac{1}{2} \log \left( 1 + \frac{p_k}{A-p_k} \right) \\
& & + \frac{1}{2} \log \left( 1 + \frac{c-p_k}{A-Mp_k-\left( c - p_k \right)}\right).
\end{eqnarray*}%
Setting $\frac{df\left( p_k \right)}{dp_k} = 0$ subject to $0 \leq p_k \leq c$, the unique solution is $p^{*}_k =
\frac{1}{M} ( A - \sqrt{A(A-cM )} )$. Thus, $p^{*}_k$ is the unique stationary point of $f \left( p_k \right)$. We can
also verify that $f\left( p^{*}_k \right) > f \left( 0 \right)$ and $f\left( p^{*}_k \right)
> f \left( c \right)$. So $\left(p^{*}_k, c-p^{*}_k \right)$ is the unique
solution to our maximization problem. We can directly solve $\frac{p_k}{A-p_k} = \frac{p_j}{A-Mp_k-p_j}$
subject to $p_k+p_j=c$ and $p_k,p_j \geq 0$. The unique solution is also $\left(p^{*}_k, c-p^{*}_k \right)$.
\qed

We now present the optimal splitting method. Theorem \ref{thm:nece} states a necessary condition for the optimal
splitting method, and Theorem \ref{thm:unique} implies there is a unique optimal splitting method. In Corollary
\ref{cor:optimal}, we formally present the optimal splitting method and the required power levels. \vspace{0.5cm}
\begin{theorem} \label{thm:nece}
Let each real user split its power into $L$ virtual users. Let $p_k$ be the power allocated to the $k$th virtual user
and $r_k = \frac{1}{2} \log$ $( 1 + \frac{p_k}{MP - M \sum_{j<k}p_j-p_k+N} )$. If $\left( p^{*}_1,...p^{*}_L \right)$
maximizes $\sum^{L}_{k=1}r_k$ and satisfies $\sum^{L}_{k=1}p^{*}_k = P$, $p_k^{*} \geq 0$ for $k=1,2...,L$, then $r_k
\left(p^{*}_1,...p^{*}_L \right) = r^{*}$, for all $k$.
\end{theorem}\vspace{0.1cm}
That is, the optimal power split must lead to equal transmission rates for all virtual users.

\vspace{0.2cm} \textit{Proof:} We use a perturbation argument. Suppose $\left( \tilde{p}_1,...,\tilde{p}_L \right)$
maximizes $\sum_{k=1}^L r_k$ and satisfies $\sum^L_{k=1}\tilde{p}_k = P$, $\tilde{p}_k \geq 0 \text{ } \forall k$ and
the resulting $r_k \left( \tilde{p}_1,...,\tilde{p}_L \right)$ is not the same for all $k$. Then we can find a pair of
virtual users $\left(k,k+1 \right)$, where virtual user $k$ and $k+1$ are decoded at the $k$th and $\left( k+1
\right)$th places respectively, and $r_k \left( \tilde{p}_1,...,\tilde{p}_L \right) \neq r_{k+1} \left(
\tilde{p}_1,...,\tilde{p}_L \right)$.

Without loss of generality, let us consider the case where $r_k\left(\tilde{p}_1,...,\tilde{p}_L \right)$ $> r_{k+1}
\left(\tilde{p}_1,...,\tilde{p}_L\right)$. By the definition of $r_k$, we have
\begin{multline}\label{eq:same} \frac{\tilde{p}_k}{MP - M \sum_{j<k}\tilde{p}_j-\tilde{p}_k+N} \\
> \frac{\tilde{p}_{k+1}}{MP - M \sum_{j<k}\tilde{p}_j-M\tilde{p}_k - \tilde{p}_{k+1}+N}. \end{multline}

We can verify that if we change $\tilde{p}_k$ to $\tilde{p}_k-\varepsilon$ and $\tilde{p}_{k+1}$ to
$\tilde{p}_{k+1}+\varepsilon$, where $\varepsilon$ is a small positive number, then the first term of \eqref{eq:same}
decreases and the second term of \eqref{eq:same} increases. Let $\varepsilon^{*}$ be the solution to \begin{multline*}
\frac{\tilde{p}_k-\varepsilon}{MP - M \sum_{j<k}\tilde{p}_j- \left(\tilde{p}_k-\varepsilon\right)+N} \\ =
\frac{\tilde{p}_{k+1}+\varepsilon}{MP - M \sum_{j<k}\tilde{p}_j- M\left(\tilde{p}_k-\varepsilon\right)- \left(
\tilde{p}_{k+1}+\varepsilon\right)+N}. \end{multline*} (The existence of $\varepsilon^*$ can be demonstrated). Let
$\hat{p}_k = \tilde{p}_k - \varepsilon^{*}$ and $\hat{p}_{k+1} = \tilde{p}_{k+1} + \varepsilon^{*}$. Notice that
$\tilde{p}_k+\tilde{p}_{k+1}= \hat{p}_k + \hat{p}_{k+1}$. Since the maximization considered in Lemma
\ref{lemma:maximization} has a unique solution, $r_k\left(\tilde{p}_1,...,\hat{p}_k,\hat{p}_{k+1},...,\tilde{p}_L
\right) + r_{k+1}\left(\tilde{p}_1,...,\hat{p}_k,\hat{p}_{k+1},...,\tilde{p}_L \right)
>
r_k\left(\tilde{p}_1,...,\tilde{p}_k,\tilde{p}_{k+1},...,\tilde{p}_L \right) +
r_{k+1}\left(\tilde{p}_1,...,\tilde{p}_k,\tilde{p}_{k+1},...,\tilde{p}_L \right) $. This contradicts our
assumption that $\left( \tilde{p}_1,...,\tilde{p}_L \right)$ maximizes \\ $\sum_{k=1}^L r_k$. Therefore, the
theorem follows. \qed

By Theorem \ref{thm:nece}, if $\left( p^{*}_1,...,p^{*}_L \right)$ maximizes $\sum^L_{k=1}r_k$ and satisfies
$\sum^L_{k=1}p^{*}_k=P$, $p^{*}_k \geq 0$ for $k=1,2...,L$, then we must have $SIR_1\left(p^{*}_1,...,p^{*}_L \right)=
...=SIR_L\left(p^{*}_1,...,p^{*}_L \right)$ where $SIR_k\left(p_1,...,p_L \right) = \frac{p_k}{MP - M \sum_{j<k}p_j-
p_k+N}$. Therefore, if we show that $SIR_k\left(p_1,...,p_L \right) = SIR_j\left(p_1,...,p_L \right) \text{ }$ for all
$ k \neq j$, has a unique solution, then there is \textit{at most} one feasible solution to the maximization problem.

\vspace{0.5cm}
\begin{theorem} \label{thm:unique}
The set of equations: $SIR_1 \left( p_1,...,p_L \right) =... = SIR_L \left( p_1,...,p_L \right)$, subject to
$\sum^L_{k=1}p_k = P $ and $p_k \geq 0$ $\forall k$, has a unique solution.
\end{theorem}\vspace{0.3cm}

\textit{Proof:} See Appendix I. \vspace{0.2cm}

\begin{corollary} \label{cor:optimal}
If a real user splits its power $P$ into $L$ virtual users, then the unique way to maximize this user's throughput is
to set $p_k = \frac{N}{M}(1+\frac{MP}{N})^{\frac{L-k}{L}}[(1+\frac{MP}{N})^{\frac{1}{L}}-1]$ for $k = 1,...,L$.
\end{corollary}\vspace{0.3cm}

\textit{Proof:} Since the constraint region $\{{\bf{p}}:\sum_{k=1}^L p_k =P \text{ and } p_k \geq 0 \text{ for }
k=1,2,..,L \}$ is a simplex and $\sum_{k=1}^L r_k$ is continuous, there exists at least one solution. We denote one
solution by $\left(\hat{p}_1,...,\hat{p}_L\right)$. By the necessary condition stated in Theorem~\ref{thm:nece},
$\left(\hat{p}_1,...,\hat{p}_L\right)$ must satisfy $r_1\left(\hat{p}_1,...,\hat{p}_L\right) = ... =
r_L\left(\hat{p}_1,...,\hat{p}_L\right)$. Moreover, by the uniqueness property stated in Theorem~\ref{thm:unique},
$\left(\hat{p}_1,...,\hat{p}_L\right)$ is the unique solution to $\max_{p_1,..,p_L} \sum_{k=1}^L r_k$ subject to
$\sum_{k=1}^Lp_k =P \text{ and } p_k \geq 0 \text{ } \forall k.$ Next, we plug $p_k$ into the expression of
$SIR_k\left(p_1,...,p_L\right)$. Let $A = 1+\frac{MP}{N}$, we are able to verify that $SIR_k\left(p_1,...,p_L\right)
=\frac{A^{\frac{1}{L}}-1}{\left(M-1\right)A^{\frac{1}{L}}+1}$, which is independent of $k$. Hence, the corollary
follows. \qed

We now examine the asymptotic behavior of the DRS scheme. We first demonstrate the interesting fact that the rate tuple
converges to the maximum equal rate point for a general power split as long as all virtual users' powers go to $0$ as
$L \rightarrow \infty$. This implies a convergence result for the optimal power split. We then analyze the rate of
convergence under the optimal power split.

\vspace{0.3cm}
\begin{theorem} \label{thm:genesplit}
Given any power split $\left( p_1,...,p_L \right)$, a sufficient condition for $\lim_{L \rightarrow \infty}
\sum_{k=1}^L r_k = \frac{1}{2M}\log (1+\frac{MP}{N})$ is $\max_{k \in \{1,...,L\}} p_k \rightarrow 0$ as $L \rightarrow
\infty$.
\end{theorem}
\vspace{0.3cm}

\textit{Proof:}
\begin{eqnarray}
& & \lim_{L \rightarrow \infty} \sum_{k=1}^L r_k \nonumber \\
& = & \lim_{L \rightarrow \infty}
\sum_{k=1}^L \frac{1}{2} \log \left( 1+ \frac{p_k}{MP - M\sum_{j<k}p_j - p_k + N} \right) \nonumber \\
& = & \lim_{L \rightarrow \infty}
\sum_{k=1}^L \frac{1}{2} \left(\frac{p_k}{MP - M\sum_{j<k}p_j - p_k + N} \right) \label{eq:limsum}\\
& \geq & \lim_{L \rightarrow \infty}
\sum_{k=1}^L \frac{1}{2} \left(\frac{p_k}{MP - M\sum_{j<k}p_j + N} \right) \\
& = & \lim_{L \rightarrow \infty}
\sum_{k=1}^L \frac{1}{2} \left(\frac{\beta_k - \beta_{k-1}}{MP + N - M\beta_{k-1}} \right) \\
& = & \frac{1}{2} \int_0^P \frac{1}{MP+N-Mx} dx \\
& = & \frac{1}{2M} \log \left( 1 + \frac{MP}{N} \right)
\end{eqnarray}where $\beta_0 = 0$, $\beta_k = \sum_{j=1}^k p_j$.

The equality in \eqref{eq:limsum} is justified as follows. Note that, $ \lim_{x \rightarrow 0}
\frac{\log(1+x)-x}{x^2} = - \frac{1}{2}$. Hence, for sufficiently small positive $\delta$, $|x| < \delta
\Rightarrow |\frac{\log(1+x)-x}{x^2}| < 1 \Rightarrow |\log(1+x) -x| \leq x^2$. Now, we examine the error
term

\begin{eqnarray}
& &\left| \sum_{k=1}^L \left[ \log \left( 1+ \frac{p_k}{MP - M\sum_{j<k}p_j - p_k + N} \right) \right. \right. \nonumber \\
& & \hspace{1cm} \left. \left. - \left(\frac{p_k}{MP - M\sum_{j<k}p_j -
p_k + N} \right)\right] \right| \nonumber \\
& \leq & \sum_{k=1}^L \left| \log \left( 1+ \frac{p_k}{MP - M\sum_{j<k}p_j - p_k + N} \right) \right. \nonumber \\
& & \hspace{1cm} \left. - \left(\frac{p_k}{MP - M\sum_{j<k}p_j -
p_k + N} \right) \right| \nonumber \\
& \leq & \sum_{k=1}^L \left(\frac{p_k}{MP - M\sum_{j<k}p_j -
p_k + N} \right)^2 \label{eq:small} \\
& \leq &\sum_{k=1}^L \left(\frac{1}{N}\right)^2 p_k^2 \label{inq:powerconv}\\
& \leq & \left(\frac{1}{N}\right)^2 \max_{k} p_k \sum_{k=1}^L p_k \\
& = & P \left(\frac{1}{N}\right)^2 \max_{k} p_k
\end{eqnarray}where inequality in~\eqref{eq:small} holds because $\frac{p_k}{MP - M\sum_{j<k}p_j - p_k + N} < \delta$ when $L
\rightarrow \infty$, and the inequality in \eqref{inq:powerconv} follows from the fact that $\frac{p_k}{MP -
M\sum_{j<k}p_j - p_k + N} \leq \frac{p_k}{N}$. Since $\max_{k} p_k \rightarrow 0$ as $L \rightarrow \infty$, the error
term goes to zero in the limit. This justifies the equality in \eqref{eq:limsum}. Using the capacity bound, we also
have $\lim_{L \rightarrow \infty} \sum_{k=1}^L r_k \leq \frac{1}{2M} \log ( 1 + \frac{MP}{N} )$. Therefore, $\lim_{L
\rightarrow \infty} \sum_{k=1}^L r_k = \frac{1}{2M} \log ( 1 + \frac{MP}{N} ).$ \qed \vspace{0.3cm}

Note that our optimal power split satisfies the sufficient condition in Theorem~\ref{thm:genesplit}. Therefore, its
convergence is assured. \vspace{0.2cm}
\begin{corollary} \label{cor:conv}
If each real user adopts the optimal splitting method specified in Corollary~\ref{cor:optimal}, then
\begin{equation*}
\lim_{L\rightarrow \infty} \sum_{k=1}^L r_k = \frac{1}{2M}\log \left(1+\frac{MP}{N}\right) \equiv R^*.
\end{equation*}
\end{corollary}
\vspace{0.5cm}

Next, we examine the rate of convergence to the maximum equal rate
point under the optimal power split. Define the error term $e[L]
\equiv R^* - \sum_{k=1}^L r_k$, we analyze how fast this error
term tends to $0$ as $L \rightarrow \infty$. We prove the
following:\footnote{Recall that $f\left[n\right] =
\Omega\left(g\left[n\right]\right)$ if there exist positive
constants $c_1$ and  $n_0$ such that $f\left[n\right] \geq c_1
g\left[n\right] $ for all $n \geq n_0$, and $f[n] = O(g[n])$ if
there are positive constants $c_2$ and $n_0$, such that $ f[n]
\leq c_2 g[n]$ for all $ n \geq n_0$. Finally, $f[n] = \Theta
\left(g[n]\right)$ if $f[n] = \Omega \left(g[n]\right)$ and $f[n]
= O \left(g[n]\right)$.}

\vspace{0.3cm}
\begin{theorem} \label{thm:speed}
$e\left[L\right] = \Theta \left(\frac{1}{L}\right)$.
\end{theorem}
\vspace{0.2cm}

\textit{Proof:} Let $A \triangleq 1+\frac{MP}{N}$.

\begin{eqnarray}
& & \lim_{L \rightarrow \infty} L e\left[L\right] \nonumber \\
& = & \lim_{L \rightarrow \infty} L \left( \frac{1}{2M}\log
\left(A\right) - \frac{L}{2} \log \left( \frac{M}{M-1 + A^{-\frac{1}{L}}}\right)\right) \nonumber \\
& = & \lim_{y \rightarrow 0} \frac{\frac{y\log\left(A\right)}{2M}-\frac{1}{2} \log
\left(\frac{M}{M-1+A^{-y}}\right)}{y^2}\\
& = & \lim_{y \rightarrow 0}
\frac{\frac{\log\left(A\right)}{2M}-\frac{1}{2}\frac{A^{-y}\log\left(A\right)}{M-1+A^{-y}}}{2y} \label{eq:l1}\\
& = & \frac{\left(M-1\right)\left(\log\left(A\right)\right)^2}{4M^2}.\label{eq:l2}
\end{eqnarray}Note that equalities in \eqref{eq:l1} and \eqref{eq:l2} can be verified by L'Hospital's rule.

Consequently, given any $\varepsilon > 0$, there exists a positive integer $n_0$ such that for all $L>n_0$,
we have $\left| L e\left[L\right] - \frac{\left(M-1\right)\left(\log\left(A\right)\right)^2}{4M^2}\right| <
\varepsilon$. This implies $\frac{\left(M-1\right)\left(\log\left(A\right)\right)^2}{4M^2} - \varepsilon < L
e\left[L\right] < \frac{\left(M-1\right)\left(\log\left(A\right)\right)^2}{4M^2} + \varepsilon$. Therefore,
we can choose a small enough $\varepsilon$ such that
$\frac{\left(M-1\right)\left(\log\left(A\right)\right)^2}{4M^2} - \varepsilon > 0$. Let $c_1 =
\frac{\left(M-1\right)\left(\log\left(A\right)\right)^2}{4M^2} - \varepsilon, c_2 =
\frac{\left(M-1\right)\left(\log\left(A\right)\right)^2}{4M^2} + \varepsilon$, we have $c_1 \leq L
e\left[L\right] \leq c_2$. This implies that there exists a positive integer $n_0$ and for all $L>n_0$, we
have $\frac{c_1}{L} \leq e\left[L\right] \leq \frac{c_2}{L}$. \qed \vspace{0.3cm}

\begin{figure}
\begin{center}
\includegraphics[width = 6.5cm, height = 6cm]{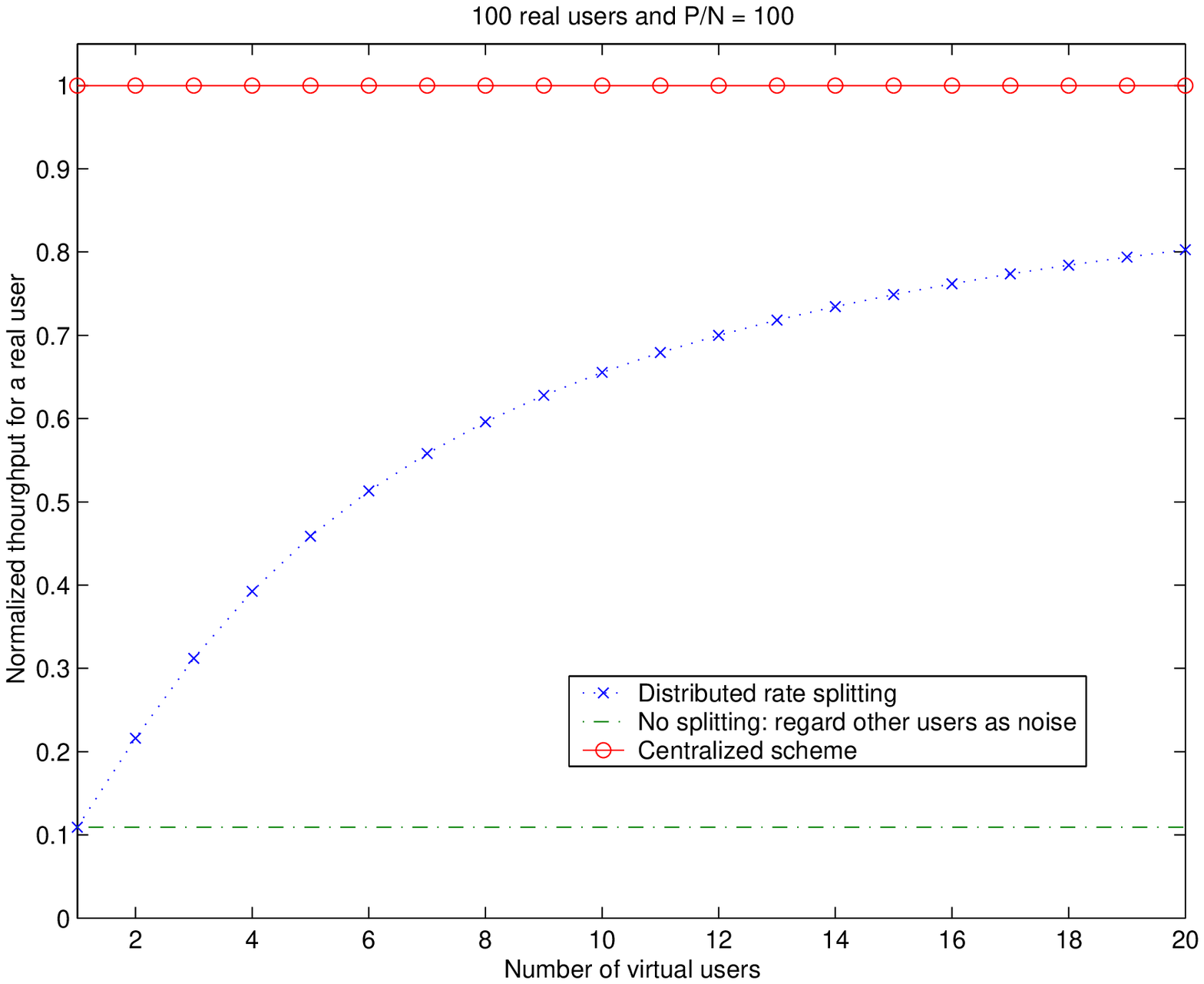}
\includegraphics[width = 6.5cm, height = 6cm]{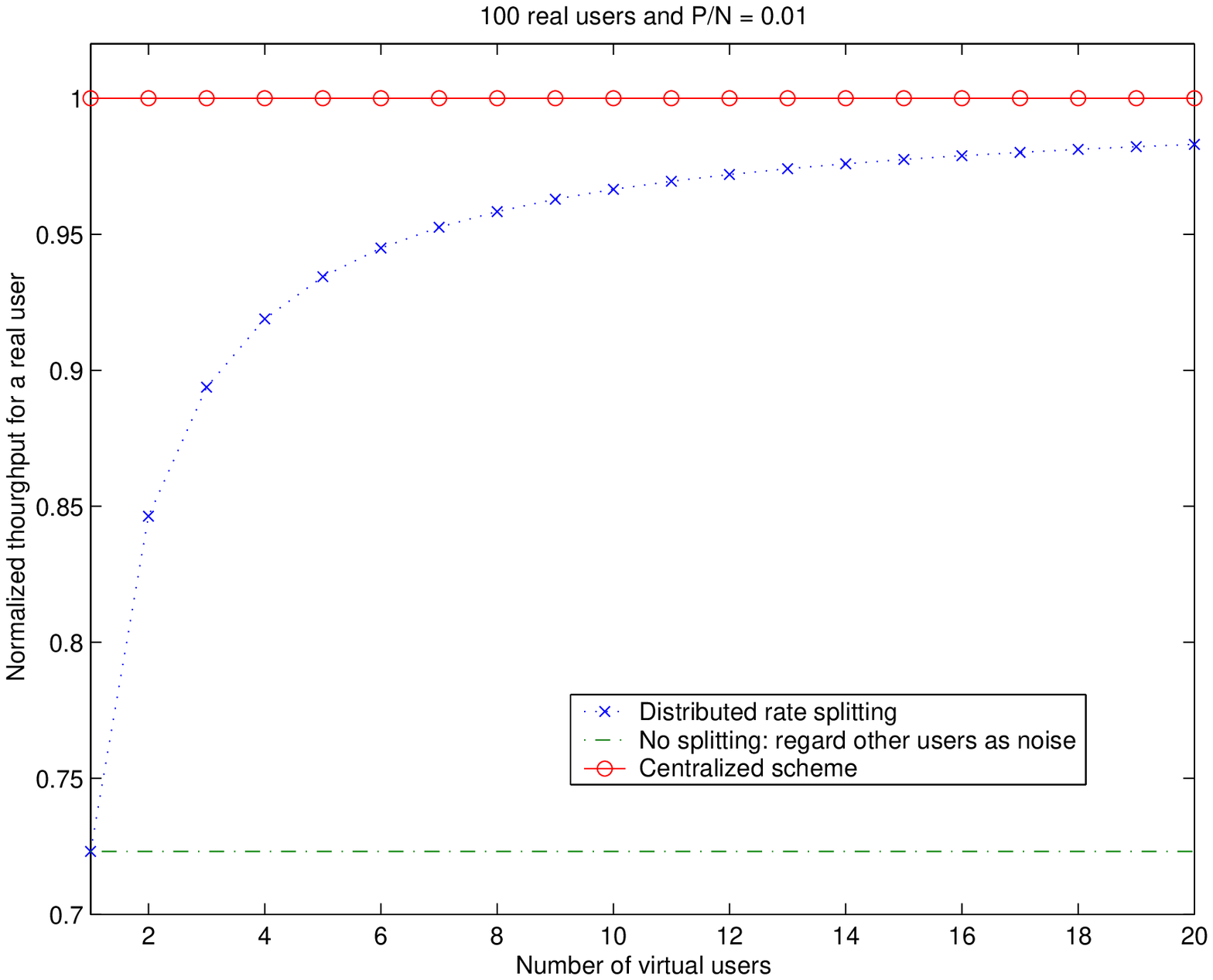}
\caption{Throughput per real user v.s. the number of virtual users per real user for both high and low SNR regimes.
Note that the scales of the vertical axes in both figures are different.} \label{fig:simulation}
\end{center}
\end{figure}

Finally, we note that all virtual users in one virtual user class can be decoded \textit{in parallel}. Thus, the
decoding delay of DRS is proportional to the number of virtual users $L$ and independent of the number of real users.
Since $L$ is controlled by the designer, DRS offers a tradeoff between the throughput of a real user and the decoding
delay. In Fig.~\ref{fig:simulation}, we present some numerical simulations illustrating the tradeoff between the number
of virtual users and the throughput for both the high and low SNR regimes. A system with $100$ real users is used in
the simulations.

\section{M-user Discrete Memoryless MAC} \label{sec:dmc}

%\subsection{Optimal Transmission Rate Distribution Function}

An $M$-user discrete memoryless MAC is defined in terms of $M$ discrete input alphabets ${\cal X}_i$, $i \in
\{1,...,M\}$, an output alphabet $\cal Y$ and a stochastic matrix $W : {\cal X}_1 \times {\cal X}_2 \times
\cdots \times {\cal X}_M \rightarrow \cal Y$ with entries $W\left(y \mid x_1,...,x_M \right)$. For any
product input distribution $P_{X_1}\cdots P_{X_M}$, let the achievable region ${\cal R}[W; P_{X_1} \cdots
P_{X_M}]$ be the set of $\mathbf{R} \in \mathbb{R}_+^M$ satisfying
\[ \sum_{i \in S} R_i
\leq I\left( X_{S};Y \mid X_{{S}^c} \right), \forall S \subseteq \{1,...,M\}
\]
where $ X_{S} \triangleq (X_i)_{i \in S}$ and $ S^c \triangleq \{1,...,M\} \setminus S$.

The capacity region of the asynchronous MAC is \cite{paper:pol83} \cite{paper:hh85}
\begin{equation}
{\cal C} = \bigcup_{P_{X_1} \cdots P_{X_M}} {\cal R}[W;P_{X_1} \cdots P_{X_M}].
\end{equation}
We fix the input product distribution $P_{X_1} \cdots P_{X_M}$ and focus on achieving the desired operating point in
${\cal R}[W; P_{X_1} \cdots P_{X_M}]$. In this section, we consider only discrete memoryless channels satisfying the
following symmetry condition:
\begin{equation} \label{eq:sym_dmc}
I \left( X_{S};Y \mid X_{S^c} \right) = I \left(X_{T};Y \mid X_{T^c} \right),
\end{equation}
$ \forall S,T \subseteq \{1,...,M\}$ such that $|S| = |T|$. Later,
in Section~\ref{sec:variation}, we consider the more general
asymmetric case.  We further assume that for $\forall S, T
\subseteq \left\{1,2,...,M\right\}$, if $S \cap T = \varnothing$,
then $I \left( X_{S};Y \right) < I \left( X_{S};Y | X_{T}
\right)$.  Under our symmetric setting, the maximum common rate
that every user can achieve is $R^* =
\frac{1}{M}I\left(X_1,...,X_M;Y\right)$.

In the Gaussian MAC, virtual users are created via a power/rate splitting mechanism. For the discrete memoryless MAC,
we adopt the random switching mechanism of~\cite{paper:rim01} where virtual users with the same input distribution as
the real users are created and the transmitted signal of a real user is determined by a random switch. We first
consider the two-user discrete memoryless MAC $(M = 2)$, and illustrate the random switching mechanism. The optimal
random switches and the asymptotic behavior of DRS under the optimal switching are presented. We then examine the
$M$-user case $(M > 2)$, and present a sufficient condition for the random switching mechanism to converge to the
information theoretic upper bound. Finally, we investigate the rate of convergence for a simple suboptimal random
switch.

\vspace{0.2cm} \subsection{Two-user Case $(M = 2)$}

Consider a two-user MAC, $W: {\cal X}_1 \times {\cal X}_2 \rightarrow {\cal Y}$. For a fixed input product
distribution $P_{X_1}P_{X_2}$, the achievable region is given by:
\begin{eqnarray*}
R_1 & \leq & I \left( X_1;Y|X_2 \right) = I \left(X_1;Y,X_2 \right)\\
R_2 & \leq &I \left(X_2;Y|X_1 \right) = I \left(X_2;Y,X_1 \right)\\
R_1 + R_2 & \leq & I \left(X_1,X_2;Y \right).
\end{eqnarray*}

Under our symmetry assumption (cf ~\eqref{eq:sym_dmc}), we have $I(X_1;Y) = I(X_2;Y)$, $I(X_1;Y,X_2) = I(X_2;Y,X_1)$,
$I(X_1;Y,X_2)> I(X_1;Y)$, and the optimal rate tuple is
\begin{equation}
\left( R^*, R^* \right) \equiv \left(\frac{1}{2} I \left(X_1,X_2;Y \right),\frac{1}{2} I \left(X_1,X_2;Y
\right)\right).
\end{equation}
\begin{figure}
\begin{center}
\includegraphics[width = 6.5cm]{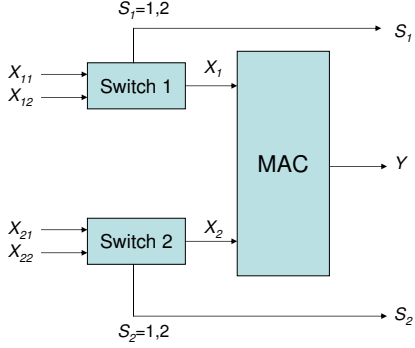}
\caption{Switches for two-user discrete memoryless MAC}\label{fig:switch}
\end{center}
\end{figure}

Let us consider the random switching mechanism for this channel. We first consider the case where each real user
generates two virtual users. Later, we consider the case where the number of virtual users per real user goes to
infinity. We split by means of two switches, as shown in Fig.~\ref{fig:switch}. Each switch has two inputs, $X_{i1} \in
{\cal X}_i$ and $X_{i2} \in {\cal X}_i$ and one output $X_i \in {\cal X}_i$. Switch $i$ is controlled by a random
variable $S_i \in \{1,2\}$ with $\mathsf{P} \left(S_i=1 \right) = \lambda$. The output is given by $ X_i = X_{i1}$ if
$S_i = 1$, and $X_i = X_{i2}$ if $S_i = 2$.

The switching random variables $\{S_1,S_2\}$ are independent of the channel inputs. We also assume that $\{S_1,S_2\}$
are available at the receiver. In practice, one would generate $S_1$ and $S_2$ at the transmitters and at the receiver,
e.g. by means of a pseudorandom sequence generator. Assign to the channel inputs $X_{11}$, $X_{12}$, $X_{21}$ and
$X_{22}$ the probability mass function $P_{X_{11},X_{12},X_{21},X_{22}}\left(x_{11},x_{12},x_{21},x_{22}\right) =
P_{X_1}(x_{11})P_{X_1}(x_{12})P_{X_2}(x_{21})P_{X_2}(x_{22}) $. Notice that $X_{i1}$ and $X_{i2}$ are independent and
each has the same probability mass function as the random variable $X_i$ for $i=1,2$.

In successive decoding for the discrete memoryless MAC, the signals of decoded virtual users are used as side
information to aid the decoding process of subsequent virtual users. The first constituent decoder observes the output
$\{Y,S_1,S_2\}$ and tries to decode $ X_{11}$ and $X_{21}$. The second constituent decoder is informed of the decision
about $\{X_{11},X_{21} \}$ made by the previous constituent decoder and tries to decode $X_{12}$ and $X_{22}$. Without
loss of generality, let us focus on real user~1.
\begin{eqnarray*}
r_{X_{11}} & = & I(X_{11};Y,S_1,S_2) \\
& = & I(X_{11};Y,S_2 | S_1 ) \\
& = & \lambda I(X_{11};Y,S_2 | S_1 = 1) \\
& & + \left(1-\lambda\right)I(X_{11};Y,S_2 |S_1 = 2)\\
& = & \lambda I(X_1;Y)
\end{eqnarray*}%
where the second equality follows from the independence between $X_{11}$ and $S_1$, and the last equality follows from
the fact that when $S_1 = 2$, $X_{11}$ is independently of the output $Y$ and $S_2$. Similarly, we have
\begin{eqnarray*}
r_{X_{12}} & = & I(X_{12};Y,S_1,S_2,X_{11},X_{21}) \\
& = & (1 - \lambda) \left[ \lambda I(X_1;Y,X_2) + (1-\lambda) I(X_1;Y) \right].
\end{eqnarray*}
It can be verified that in the two-user discrete memoryless MAC, both real users' throughput can be strictly increased
by splitting their inputs via a random switch, relative to the case where they do not split, (i.e. $r_{X_{11}} +
r_{X_{12}} > I(X_1;Y)$ for $\lambda \in (0,1)$). Next, we show that by generating more virtual users, the throughput of
each real user increases further.

\vspace{0.2cm}
\begin{lemma}
For $M = 2$, consider a distributed rate splitting scheme with $L$ virtual users per real user. The random switch for
user~$i$ is controlled by $S_i$, where $\mathsf{P}(S_i = k) = \lambda_k$ for $k = 1,...,L$. It is possible to strictly
increase the throughput via an $(L+1)$ virtual user system by splitting the $L$th virtual user into two virtual users.
\end{lemma}
\vspace{0.2cm}

\textit{Proof:} Without loss of generality, we consider user~$1$. For the $k$th virtual user, we have
\begin{eqnarray}
& & r_{X_{1k}} \nonumber\\
& = & I\left(X_{1k};Y,S_1,S_2,X_{11},X_{21},...,X_{1(k-1)},X_{2(k-1)}\right) \nonumber \\
& = & I \left(X_{1k};Y,S_2,X_{11},X_{21},..., \right. \nonumber \\
& & \hspace{2cm} \left. X_{1(k-1)},X_{2(k-1)} \mid S_1 \right) \label{eq:leind}\\
& = & \lambda_k I \left(X_{1k};Y,S_2,X_{11},X_{21},...,\right. \nonumber \\
& & \hspace{2cm} \left. X_{1(k-1)},X_{2(k-1)} \mid S_1 = k \right) \nonumber \\
& & + \left(1-\lambda_k \right) I \left(X_{1k};Y,S_2,X_{11},X_{21},..., \right. \nonumber \\
& & \hspace{2cm} \left. X_{1(k-1)},X_{2(k-1)} \mid S_1 \neq k \right) \nonumber \\
& = & \lambda_k I \left(X_{1k};Y,S_2,X_{11},X_{21},..., \right. \nonumber \\
& & \hspace{2cm} \left. X_{1(k-1)},X_{2(k-1)} \mid S_1 = k \right) \label{eq:leind2} \\
& = & \lambda_k \left[ \left(\sum_{j<k} \lambda_j \right)I \left(X_{1k};Y,X_{11},X_{21},..., \nonumber \right. \right. \nonumber \\
& & \hspace{2cm} \left.  X_{1(k-1)},X_{2(k-1)} \mid S_1 = k, S_2 < k \right) \nonumber \\
& & + \left( 1 - \sum_{j<k} \lambda_j \right) I \left(X_{1k};Y,X_{11},X_{21},..., \right. \nonumber \\
& & \hspace{2cm} \left.  X_{1(k-1)},X_{2(k-1)} \mid S_1 = k, S_2
\geq k \right) \Big] \nonumber \\
& = & \lambda_k \left[ \left(\sum_{j<k} \lambda_j \right)I \left(X_1;Y,X_2 \right)  \right. \nonumber \\
& & \hspace{2cm} + \left. \left( 1 - \sum_{j<k} \lambda_j \right) I\left(X_1;Y \right) \right], \label{eq:lelast}
\end{eqnarray}%
where equality in \eqref{eq:leind} is due to the independence between $X_{1k}$ and $S_1$, and equality
in~\eqref{eq:leind2} follows from the fact that when $S_1 \neq k$, $X_{1k}$ is independent of the output $Y$ and all
the other random variables. Finally, equality in~\eqref{eq:lelast} holds because when $S_2 < k$, one of the random
variables $X_{21},...,X_{2(k-1)}$ is the switch output, and when $S_2 \geq k$, none of them is the switch output.
Therefore, \[r_{X_{1L}}= \lambda_L \left[ \left(1-\lambda_L \right) I \left(X_1;Y,X_2 \right) \right. +  \left.
\lambda_L I \left(X_1;Y \right) \right]. \]

Now let us split $r_{X_{1L}}$ into $r_{X_{1L}^1}$ and $r_{X_{1L}^2}$ by using a switch controlled by a binary random
variable $S'$ with $\mathsf{P}(S' = 0) = \alpha$. We have
\begin{eqnarray*}
r_{X_{1L}^1} & = & \alpha \lambda_L \left( \left(1-\lambda_L \right)
I\left(X_1;Y,X_2\right) + \lambda_L I \left( X_1;Y \right)\right),\\
r_{X_{1L}^2} & = & \overline{\alpha} \lambda_L \left( \left(1- \overline{\alpha} \lambda_L \right)I \left(X_1;Y,X_2
\right) + \overline{\alpha} \lambda_L I \left(X_1;Y \right) \right),
\end{eqnarray*}
where $\bar{\alpha} = 1-\alpha$. Hence,
\begin{eqnarray*}
& & r_{X_{1L}^1}+r_{X_{1L}^2}-r_{X_{1L}} \\
& = & \alpha \lambda_L \left( \left(1-\lambda_L \right) I \left(X_1;Y,X_2 \right) + \lambda_L I \left(X_1;Y
\right)\right) \nonumber \\
& & + \overline{\alpha} \lambda_L \left( \left(1- \overline{\alpha} \lambda_L \right)I \left(X_1;Y,X_2 \right) +
\overline{\alpha}
\lambda_L I \left(X_1;Y \right) \right)\\
& & - \lambda_L \left[ \left(1-\lambda_L
\right) I\left(X_1;Y,X_2\right) + \lambda_L I \left(X_1;Y \right) \right] \\
& = & \alpha \overline{\alpha} \lambda_L^2 \left[I \left(X_1;Y,X_2 \right) - I \left(X_1;Y \right) \right] \\
& \geq & 0
\end{eqnarray*}
with strict inequality if $\alpha \in (0,1)$. \qed

Before we examine the asymptotic behavior of the DRS scheme for $M = 2$, we first solve the problem of how to find the
optimal switches for a fixed number of virtual users per real user.

\vspace{0.2cm}
\begin{theorem}
For $ M = 2$, if a real user has $L$ virtual users, then the optimal random variable to control the switch
for user $i$ is $S_i \in \{1,...,L\}$ with $\mathsf{P}(S_i=k) = \frac{1}{L}$ for $k = 1,...,L$ and $i=1,2$.
\end{theorem}
\vspace{0.2cm}

\textit{Proof:} We use a perturbation argument. Suppose the random variables $S_1,S_2 \in \{1,...,L\}$ with
$\mathsf{P}(S_1=k)=\mathsf{P}(S_2 = k) = \lambda_k$ maximize $R_{X_1}$, where $\lambda_k \geq 0$ and $\sum_{k=1}^L
\lambda_k = 1$. Moreover, suppose there exists $\lambda_k$ such that $\lambda_k \neq \frac{1}{L}$. Let $\lambda_k$ be
the first element which is not equal to $\frac{1}{L}$. We consider the pair $(\lambda_k, \lambda_{k+1})$. We have
\begin{multline*}
r_{1k}  =  \lambda_k \left(\frac{k-1}{L} I(X_1;Y,X_2) \right.  \\
 + \left. \left(1-\frac{k-1}{L} \right)I(X_1;Y) \right), \end{multline*}
\begin{multline*} r_{1(k+1)}  =  \lambda_{k+1} \left[\left(\frac{k-1}{L}+\lambda_k \right)I(X_1;Y,X_2) \right.  \\
 \left. + \left(1-\frac{k-1}{L}-\lambda_k \right)I(X_1;Y) \right].
\end{multline*}

Therefore,
\begin{eqnarray*}& & r_{1k}+r_{1(k+1)} \\
& = & \left[\left(\lambda_k+\lambda_{k+1}\right) \frac{k-1}{L} +
\lambda_k \lambda_{k+1} \right]I(X_1;Y,X_2) \\
& & + \left[\left(\lambda_k + \lambda_{k+1}\right)\frac{L-k+1}{L} - \lambda_k \lambda_{k+1} \right] I(X_1;Y).
\end{eqnarray*}

First, we consider the case where $\lambda_k > \lambda_{k+1}$. We let $\hat{\lambda}_k= \lambda_k-\varepsilon$ and
$\hat{\lambda}_{k+1}= \lambda_{k+1}+\varepsilon$ for $\varepsilon < \frac{\lambda_k - \lambda_{k+1}}{2}$. We have
\begin{multline*}\hat{r}_{1k} + \hat{r}_{1(k+1)}\\
=\left[ \left(\lambda_k +\lambda_{k+1} \right)\frac{k-1}{L} + \left(\lambda_k-\varepsilon \right)
\left(\lambda_{k+1}+\varepsilon \right)\right]I(X_1;Y,X_2) \\
 +  \left[ \left(\lambda_k + \lambda_{k+1} \right)\frac{L-k+1}{L} - \left(\lambda_k-\varepsilon \right) \left(\lambda_{k+1}+\varepsilon \right) \right]
I(X_1;Y). \end{multline*}

Thus, $ \hat{r}_{1k}$ + $\hat{r}_{1(k+1)}$ - $(r_{1k}$ + $r_{1(k+1)})$ = $[ (\lambda_k $ + $ \lambda_{k+1} )$
$\varepsilon$ -$\varepsilon^2$] $(I(X_1;Y,X_2)$-$I(X_1;Y) ).$ Notice that the second term of the R.H.S. expression is
positive and there exists $\varepsilon$ such that $((\lambda_k$ + $\lambda_{k+1})$ $\varepsilon$-$\varepsilon^2) > 0$.
This, however, contradicts our assumption that the random variables $S_1,S_2 \in \{1,...,L\}$ with $\mathsf{P}(S_1=k) =
\mathsf{P}(S_2=k) = \lambda_k$, maximize $R_{X_1}$. We obtain similar contradictions for the case where $\lambda_k \leq
\lambda_{k+1}$. \qed

We now examine the asymptotic behavior of the DRS scheme in the two-user discrete memoryless MAC. \vspace{0.2cm}
\begin{theorem} \label{thm:twoconv}
For $M = 2$, if both real users adopt $\mathsf{P}\left(S_i = k\right) = \frac{1}{L}$ for $k=1,...,L$ and $i = 1,2$,
then $\lim_{L \rightarrow \infty} \sum_{k=1}^L r_{X_{ik}} = \frac{1}{2} I \left( X_1,X_2;Y \right) = R^*$ for $i =
1,2$. Moreover, if we define the error term $e[L] \equiv R^* - \sum_{k=1}^L r_{X_{ik}}$, then $e[L] = \Theta
(\frac{1}{L} )$ for $i = 1,2$.
\end{theorem}
\vspace{0.3cm}

\textit{Proof:} Without loss of generality, we consider real user~$1$. Since
\begin{equation*}
r_{X_{1k}}=\frac{1}{L} \left( \frac{k-1}{L}I(X_1;Y,X_2) + \left(1-\frac{k-1}{L}\right) I(X_1;Y)\right),
\end{equation*}
we have \begin{eqnarray*} & & \lim_{L \rightarrow \infty} R_{X_1} \nonumber \\
& = & \lim_{L \rightarrow \infty} \left[\frac{1}{L} \left(1 + \frac{1}{L} + ... + \frac{L-1}{L} \right)I(X_1;Y)  \right. \nonumber \\
& &  \left.   + \frac{1}{L}\left(\frac{L-1}{L} +
... + \frac{1}{L} \right)I(X_1;Y,X_2) \right] \\
& = & \lim_{L \rightarrow \infty}\left[\frac{L+1}{2L}I(X_1;Y) + \frac{L-1}{2L}I(X_1;YX_2)\right] \\
& = & \frac{1}{2} \left[I(X_1;Y) +I(X_1;Y,X_2) \right] \\
& = & \frac{1}{2}I(X_1,X_2;Y),
\end{eqnarray*}%
where the last equality follows from the fact that $I(X_1;Y,X_2) = I(X_2;Y,X_1)$ in the symmetric setting.

Next, we examine the rate of convergence. We have the error term $e[L] = \frac{1}{2}I(X_1;Y) + \frac{1}{2}I(X_1;Y,X_2)
- (\frac{L+1}{2L}I(X_1;Y) + \frac{L-1}{2L}I(X_1;Y,X_2))$. Thus,
\[
e[L] = \frac{1}{2L}\left( I(X_1;Y,X_2) - I(X_1;Y) \right),
\]
which implies
\[
\lim_{L \rightarrow \infty} Le[L] = \frac{1}{2}\left( I(X_1;Y,X_2) - I(X_1;Y) \right) \triangleq \frac{c}{2}.
\]
For any $\varepsilon>0$, there exists an $n_0$ such that for all $L>n_0$, $|Le[L] - \frac{c}{2}|<\varepsilon$. Hence,
$\frac{c}{2} - \varepsilon < Le[L] <\frac{c}{2}+\varepsilon $. We can choose $\varepsilon$ small enough such that
$\frac{c}{2} - \varepsilon > 0$. This implies $e[L] = \Theta (\frac{1}{L} )$. \qed

%================================================================
% M-user Case
%================================================================
\subsection{$M$-user Case $(M > 2)$} There are $M$ real users and each real user creates $L$ virtual users. We have
$M$ switches $\left(S_1,...,S_M \right)$ to do the splitting with probabilities $\mathsf{P}(S_i = k) = \lambda_k$ for
$i = 1,...,M$ and $ k =1,...,L$. We assume the receiver also knows $\left(S_1,...,S_M \right)$. This may require common
randomness to exist between all transmitters and the receiver. Due to symmetry, we focus on one user, say user~$1$. For
$k = 1,...,L$, we set
\begin{eqnarray}
& & r_{X_{1k}} \nonumber \\
& = & I(X_{1k};Y,\textbf{S}_1^M,\textbf{X}_{11}^{1(k-1)},...,\textbf{X}_{M1}^{M(k-1)} ) \label{eq:simplify}\\
& = & I(X_{1k};Y,\textbf{S}_2^M,\textbf{X}_{11}^{1(k-1)},...,\textbf{X}_{M1}^{M(k-1)}\mid S_1 ) \label{eq:mind} \\
& = & \lambda_k I(X_{1k};Y,\textbf{S}_2^M,\textbf{X}_{11}^{1(k-1)},...,\textbf{X}_{M1}^{M(k-1)}
\mid S_1 = k) \nonumber \\
& & +  \left(1-\lambda_k \right) I(X_{1k};Y,\textbf{S}_2^M, \textbf{X}_{11}^{1(k-1)},...,  \nonumber \\
& & \textbf{X}_{M1}^{M(k-1)}
\mid S_1 \neq k) \nonumber \\
& = & \lambda_k I(X_{1k};Y,\textbf{S}_2^M,\textbf{X}_{11}^{1(k-1)},...,\textbf{X}_{M1}^{M(k-1)}
\mid S_1 = k) \label{eq:mind2} \\
& = & \lambda_k I(X_{1};Y,\textbf{X}_{11}^{1(k-1)},...,\textbf{X}_{M1}^{M(k-1)}
| S_1 = k, \textbf{S}_2^M) \nonumber \\
& = & \sum_{s_2,...,s_M} \mathsf{P}\left(S_2 = s_2,...,S_M = s_M\right)  \lambda_k
I(X_{1};Y,\textbf{X}_{11}^{1(k-1)}, \nonumber \\
& & \hspace{1cm} ..., \textbf{X}_{M1}^{M(k-1)}
| S_1 = k, S_2 = s_2,...,S_M = s_M) \nonumber \\
& = & \lambda_k \left[ \left(1- \sum_{j<k} \lambda_j \right)^{M-1}
I(X_1;Y) \right. \nonumber \\
& & + \sum_{i=1}^{M-1} {M-1 \choose i} \left(\sum_{j<k}\lambda_j \right)^i  \cdot \nonumber \\
& & \left. \left(1-\sum_{j<k}\lambda_j \right)^{M-1-i} I(X_1;Y,X_2,...,X_{i+1}) \right], \label{eq:mlast}
\end{eqnarray}%
where $\textbf{S}_i^j \triangleq \{S_i,...,S_j \}$ and $\textbf{X}_{i1}^{ik} \triangleq \{X_{i1},...,X_{ik}  \}$.
Equality in~\eqref{eq:mind} is due to the independence between $X_{1k}$ and $S_1$. Equality in~\eqref{eq:mind2} holds
because when $S_1 \neq k$, $X_{1k}$ is independent of the output $Y$ and all the other random variables. The first term
in~\eqref{eq:mlast} follows from $\mathsf{P}(S_2 \geq k,...,S_M \geq k )$ = $(1- \sum_{j<k} \lambda_j )^{M-1}$. The
second summation term in~\eqref{eq:mlast} follows from the fact that the probability of $i$ switching random variables
among $(S_2,...,S_M)$ having values less than $k$ is ${M-1 \choose i} (\sum_{j<k}\lambda_j )^i (1-\sum_{j<k}\lambda_j
)^{M-1-i}$. It can be verified that real user~$1$ with $L$ virtual users can strictly increase its total throughput via
an $L+1$ virtual user system.

\vspace{0.1cm} In order to maximize the total throughput of real user~1 for fixed $L$, we need to find the optimal
$\left( \lambda_1^*,...,\lambda_L^* \right)$ to maximize $\sum_{k=1}^L r_{X_{1k}}$. This is a non-convex optimization
problem and appears to be difficult. We are able to verify that for the general $M$-user case (unlike the two-user
case), random switches with a uniform distribution are in general suboptimal. Nevertheless, it is possible to
generalize the asymptotic result of Theorem~\ref{thm:twoconv}. We first demonstrate the fact that the convergence
result holds for a general switch controlled by $S_i$, where $\mathsf{P} \left(S_i = k \right) = \lambda_k$ for
$k=1,...,L$, as long as $\max_{k \in \{1,...,L\}} \lambda_k \rightarrow 0$ as $L \rightarrow \infty$. We then analyze
the rate of convergence for a particular suboptimal switch, the uniform switch.

\vspace{0.3cm}

\begin{theorem} \label{thm:mconv}
For a general random switch controlled by $S_i$, where $\mathsf{P} (S_i = k) = \lambda_k$ for $k = 1,...,L$,
a sufficient condition for $\lim_{L \rightarrow \infty} \sum_{k=1}^L r_{X_{ik}} = \frac{1}{M}
I\left(X_1,...,X_M;Y \right)$ is $\max_{k \in \{1,...,L\}} \lambda_k \rightarrow 0$ as $L \rightarrow \infty$
for $i = 1,...,M$.
\end{theorem}
\vspace{0.3cm}

\textit{Proof:} Without loss of generality, let us examine real user~$1$.
\begin{eqnarray*} & & \lim_{L \rightarrow \infty} R_{X_1} \\ & = & \lim_{L \rightarrow
\infty} \sum_{k=1}^L r_{X_{1k}} \\
& = & \lim_{L \rightarrow \infty}\sum_{k=1}^L \lambda_k \left[ \left(1-
\sum_{j<k} \lambda_j \right)^{M-1} I(X_1;Y) \right.\\
& & + \sum_{i=1}^{M-1} {M-1 \choose i} \left. \left(\sum_{j<k}\lambda_j \right)^i \cdot \right. \\
& & \hspace{1cm} \left. \left(1-\sum_{j<k}\lambda_j \right)^{M-1-i} I(X_1;Y,X_2,...,X_{i+1}) \right] \\
& = & \lim_{L \rightarrow \infty}\sum_{i=0}^{M-1}c_i I_i,
\end{eqnarray*}%
where
\begin{eqnarray*}
I_0 & = & I(X_1;Y)\\
I_i & = & I(X_1;Y,X_2,...,X_{i+1}) \\
c_0 & = & \sum_{k=1}^L \lambda_k \left( 1 -
\sum_{j<k}\lambda_j\right)^{M-1} \\
c_i & = & \sum_{k=1}^L \lambda_k {M-1 \choose i} \left( \sum_{j<k}\lambda_j \right)^i \left(1 - \sum_{j<k}\lambda_j
\right)^{M-1-i},
\end{eqnarray*}
for $i \geq 1$.

It is sufficient to prove $\lim_{L \rightarrow \infty} c_i = \frac{1}{M}$ for all $i$. For $i \geq 1$,
\begin{eqnarray*} & & \lim_{L \rightarrow \infty} c_i \\& = & \lim_{L \rightarrow \infty} \sum_{k=1}^L \lambda_k {M-1
\choose i} \left( \sum_{j<k}\lambda_j
\right)^i \left(1 - \sum_{j<k}\lambda_j \right)^{M-1-i} \\
& = & \lim_{L \rightarrow \infty} \sum_{k=1}^L \left(\beta_k -
\beta_{k-1}\right) {M-1 \choose i} \beta_{k-1}^i \left(1 - \beta_{k-1} \right)^{M-1-i} \\
& = & {M-1 \choose i} \int_0^1 x^i \left(1-x \right)^{M-1-i} dx \\
& = & {M-1 \choose i} B\left(i+1,M-i \right) \\
& = & \frac{(M-1)!}{i!(M-i-1)!} \frac{i!(M-i-1)!}{M!} \\
& = & \frac{1}{M},
\end{eqnarray*}%
where $\beta_0 = 0$, $\beta_k = \sum_{j=1}^k \lambda_k$ and $B(m,n) = \frac{(m-1)!(n-1)!}{(m+n-1)!}$ is the beta
function. The term $c_0$ can be shown to converge to $\frac{1}{M}$ as $L \rightarrow \infty$ by a similar argument.
\qed

Next, we analyze a particular suboptimal switch, the uniform switch. Since the uniform switch satisfies the sufficient
condition in Theorem~\ref{thm:mconv}, the convergence result holds. The next lemma presents its rate of convergence.
\vspace{0.3cm}
\begin{lemma}
Consider an $M$-user discrete memoryless MAC. Let each real user have $L$ virtual users and each switch be controlled
by an i.i.d. random variable $S_i \in \{1,2,...,L\}$ with $\mathsf{P}(S_i = k) = \frac{1}{L}$, $k=1,...,L$. Define the
error term $e[L] \equiv  \frac{1}{M}I(X_1,X_2,...,X_M;Y) - \sum_{k=1}^L r_{X_{ik}} $. Then $e[L] = O ( \frac{1}{L} )$
for all~$i$.
\end{lemma} \vspace{0.2cm}

\textit{Proof:} In the uniform switch setting, $\mathsf{P}(S_i = k) = \frac{1}{L}$ for $i = 1,...,M$ and $k = 1,...,L$.
Without loss of generality, we examine the total throughput of real user~$1$.
\begin{multline*}
R_{X_1} =  \sum_{k=1}^L r_{X_{1k}} \\
=  \frac{1}{L} I(X_1;Y) + \sum_{k=2}^L \frac{1}{L} \left[ \sum_{i=0}^{M-1} {M-1 \choose i} \left(\frac{k-1}{L} \right)^i \cdot \right. \\
 \left. \left(1-\frac{k-1}{L} \right)^{M-1-i}I(X_1;Y,X_2,...,X_{i+1}) \right],
\end{multline*}where $I(X_1;Y,X_2,...,X_{i+1}) = I(X_1;Y)$ for $i = 0$.

We denote $R_{X_1} \triangleq c_0I_0 + ... + c_{M-1}I_{M-1}$, where $I_i$ is defined in the proof of
Theorem~\ref{thm:mconv} for $i \geq 0$, and
\begin{eqnarray*}
c_0 & = & \frac{1}{L} \sum_{k=1}^L \left( 1 -
\frac{k-1}{L}\right)^{M-1}\\
c_i & = & \sum_{k=1}^L \frac{1}{L} {M-1 \choose i} \left(\frac{k-1}{L} \right)^i \left(1-\frac{k-1}{L}\right)^{M-1-i}.
\end{eqnarray*}for $i \geq 1$. Therefore, the error term can be calculated as follows,
\begin{eqnarray*}
e[L] & = & \left| \frac{1}{M} I(X_1,X_2,...,X_M;Y) - \sum_{i=0}^{M-1}c_i
I_i \right|\\
& = & \left| \sum_{i=0}^{M-1} \left(\frac{1}{M} - c_i \right) I_i \right| \\
& \leq & \sum_{i=0}^{M-1} \left| \frac{1}{M} - c_i \right| I_i \\
& \leq & M \left[\max_i \left(\left| \frac{1}{M} - c_i \right| I_i \right) \right].
\end{eqnarray*}

Note that $x^i \left(1-x \right)^{M-1-i}$ is maximized at $x = \frac{i}{M-1}$. For $i \geq 1$, it can be verified that
 \begin{eqnarray*}
& & \frac{1}{L} \sum_{l = 0}^{L-1} \left(\frac{l}{L} \right)^i \left(1-\frac{l}{L} \right)^{M-1-i} \\ & & +
\left(\frac{i}{M-1}\right)^i \left(1-\frac{i}{M-1}\right)^{M-1-i} \frac{1}{L} \\ & & \geq \int_0^1 x^i \left(1-x
\right)^{M-1-i} dx,
\end{eqnarray*}

\begin{eqnarray*} & & \frac{1}{L} \sum_{l = 0}^{L-1} \left(\frac{l}{L} \right)^i \left(1-\frac{l}{L}\right)^{M-1-i} \\
& & - \left(\frac{i}{M-1}\right)^i \left(1-\frac{i}{M-1}\right)^{M-1-i} \frac{1}{L} \\
& & \leq \int_0^1 x^i \left(1-x \right)^{M-1-i} dx.
\end{eqnarray*}
Multiplying both sides of the above two inequalities by ${M-1 \choose i}$, we have
\begin{equation*}
\left|\frac{1}{M} - c_i \right| \leq {M-1 \choose i} \left(\frac{i}{M-1}\right)^i \left(1-\frac{i}{M-1}\right)^{M-1-i}
\frac{1}{L}.
\end{equation*}
Therefore,
\begin{eqnarray*}
& & \max_i \left( \left| \frac{1}{M} - c_i \right|I_i\right) \\
& \leq & \max_i \left[{M-1 \choose i}\left(\frac{i}{M-1}\right)^i \left(1-\frac{i}{M-1}\right)^{M-1-i}I_i\right]
\frac{1}{L} \\
& \equiv & \alpha \frac{1}{L},
\end{eqnarray*}
so $e[L] \leq M \alpha \frac{1}{L}$, where $\alpha > 0$. For the term $c_0$, a similar argument can be used
to show that $|c_0 - \frac{1}{M}| \leq \frac{1}{L}$. Therefore, $e[L] = O \left( \frac{1}{L} \right)$. \qed

\section{Variations of Distributed Rate Splitting} \label{sec:variation}

In Section \ref{sec:gau} and Section \ref{sec:dmc}, we imposed two
symmetry constraints. The first is that the capacity region for
the Gaussian MAC and the achievable rate region for the discrete
memoryless MAC are symmetric. The second is that users generate
the same number of virtual users. In this section, we describe two
variations of DRS. The first variation is presented in
Section~\ref{sec:unequalpower}, where we relax the symmetric
region constraint. In this case, we show that as the number of
virtual user per real user tends to infinity, the rate tuple
achieved under DRS approaches a point on the dominant face. The
second variation is presented in Section~\ref{sec:differentvuser},
where each real user may generate a different number of virtual
users. The main advantage of this variation is that it can
accommodate different user rate requirements in a distributed
fashion.

\subsection{Asymmetric Capacity/Achievable Rate Region}\label{sec:unequalpower}

\vspace{0.2cm}

\subsubsection{$M$-user Gaussian MAC}

In this section, we consider the case where real users in a Gaussian MAC may have different transmission powers (i.e.
the capacity region may not be symmetric). We assume that user~$i$ has transmission power $P_i$ and the power vector
$\left( P_1,..., P_M \right)$ is known to all users. We also assume that all real users split their powers into $L$
virtual users according to the common power splitting rule defined by the vector $(\gamma_1,\gamma_2,...,\gamma_L)$,
where $\gamma_k > 0$ $ \forall k$ and $\sum_{k=1}^L \gamma_k = 1$. The power vector for the virtual users generated by
user~$i$ is $\left(\gamma_1 P_i, ..., \gamma_L P_i \right)$ for $i = 1,..., M$.

\vspace{0.5cm}

\begin{lemma}
For any real user with $L$ virtual users, the unique way to maximize this user's throughput is to set \\ \vspace{0.05cm}\\
$\gamma_k = \frac{N}{\sum_{j=1}^M P_j}\left(1+\frac{\sum_{j=1}^M P_j}{N}\right)^{\frac{L-k}{L}}
\left[\left(1+\frac{\sum_{j=1}^M P_j}{N}\right)^{\frac{1}{L}}-1\right]$  \\\vspace{0.05cm} \\for $k = 1,...,L$.
Moreover, if all real users adopt this power allocation rule, then
\[
\lim_{L \rightarrow \infty} \sum_{k=1}^L r_{ik} = \frac{P_i}{2\sum_{j=1}^M P_j} \log \left(1 + \frac{\sum_{j=1}^M
P_j}{N} \right),
\]for $i = 1,...,M$.
\end{lemma}

\vspace{0.5cm}

\textit{Proof:} By replacing $p_k$ by $\gamma_k P_i$ for $k = 1,...,L$, we can use arguments similar to those in
Section \ref{sec:gau} to prove the following:
\begin{enumerate}
    \item Given a DRS scheme with $L$ virtual users per real user, it is possible to strictly increase the throughput via an
    $(L+1)$ virtual user system.
    \item Under the optimal power split, all virtual users generated by real user~$i$ must have the same rate for $i =
    1,...,M$. (Virtual users generated by different real users may have different rates.)
    \item For any real user with $L$ virtual users, the unique way to maximize this user's throughput is to set \\ \\
     $\gamma_k =\frac{N}{\sum_{i=1}^M P_i}\left(1+\frac{\sum_{i=1}^M P_i}{N}\right)^{\frac{L-k}{L}} \left[\left(1+\frac{\sum_{i=1}^M
P_i}{N}\right)^{\frac{1}{L}}-1\right]$ \\ \\for $k = 1,...,L$.
    \item If all real users adopt this power allocation rule,
then
\[
\lim_{L \rightarrow \infty} \sum_{k=1}^L r_{ik} = \frac{P_i}{2\sum_{j=1}^M P_i} \log \left(1 + \frac{\sum_{i=1}^M
P_i}{N} \right),
\]for $i = 1,...,M$
\end{enumerate} \qed

\begin{figure}
\begin{center}
\includegraphics[width = 6.5cm]{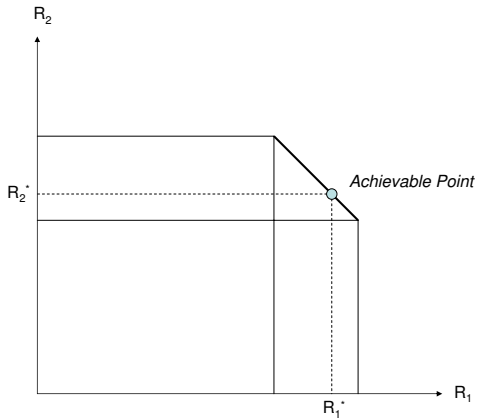}
\caption{Achievable point for a two-user Gaussian MAC where user~$1$ has higher transmission power. The transmission
powers and the achievable rates satisfy $\frac{R_1^*}{R_2^*} = \frac{P_1}{P_2}$.} \label{fig:asy_reg}
\end{center}
\end{figure}

We illustrate this achievable point on the dominant face for a two-user Gaussian MAC in Fig.~\ref{fig:asy_reg}.

\subsubsection{$M$-user Discrete Memoryless MAC}

In Section~\ref{sec:dmc}, we considered the symmetric setting (cf \eqref{eq:sym_dmc}):
\begin{equation*}
I \left( X_{S};Y \mid X_{S^c} \right) = I \left(X_{T};Y \mid X_{T^c} \right),
\end{equation*}
$ \forall S,T \subseteq \{1,...,M\}$ such that $|S| = |T|$. In this section, we relax this constraint and consider an
asymmetric achievable region. We require only that for $\forall S, T \subseteq \left\{1,2,...,M\right\}$, if $S \cap T
= \varnothing$, then $I \left( X_{S};Y \right) < I \left( X_{S};Y | X_{T} \right)$. The $M$ switches
$\left(S_1,...,S_M\right)$ have probabilities $\mathsf{P}\left(S_i = k\right) = \lambda_k$ for $i = 1,...,M$ and $k =
1,...,L$.

\vspace{0.3cm}

\begin{lemma}
Consider a general random switch controlled by $S_i$, where $\mathsf{P}\left(S_i = k\right) = \lambda_k$, $k =
1,...,L$. Define $-i \equiv \left\{1,...,M\right\} \backslash \left\{ i \right\}$. If $\max_{k \in \{1,...,L\}}
\lambda_k \rightarrow 0$ as $L \rightarrow \infty$, then
\begin{eqnarray*}
& & \lim_{L \rightarrow \infty} \sum_{k=1}^{L} r_{X_{ik}} \\
&=& \frac{1}{M} I \left(X_i;Y\right) + \sum_{j=1}^{M-1}\frac{1}{M} \frac{1}{{M-1 \choose j}}
\sum_{\substack{\varnothing \subset S \subseteq -i \\ |S| = j}} I\left(X_i;Y,X_S\right),
\end{eqnarray*} for $i = 1,...,M$.
\end{lemma}

\vspace{0.3cm}

\textit{Proof:} We can replace ${M-1 \choose i}I\left(X_1;Y,X_2,...,X_{i+1}\right)$ by $\sum_{\varnothing \subset S
\subseteq -i,|S| = j} I\left(X_i;Y,X_S\right)$ and use arguments similar to those for Theorem~\ref{thm:mconv} to prove
the above lemma. \qed

\subsection{Unequal Number of Virtual Users}\label{sec:differentvuser}

\subsubsection{$M$-user Gaussian MAC}\label{sec:adrsgau}

In this section, we retain the assumption that every user has the same transmission power $P$, but we do not require
all real users to create the same number of virtual users. That is, user~$i$ and user~$j$ create $L_i$ and $L_j$
virtual users independently, where $L_i$ may not be equal to $L_j$. The signal transmitted by a real user is the
superposition of all its virtual users' signals. We also assume in this section that user~$i$ transmits the number
$L_i$ in a header message to the receiver. The receiver receives the sum of $\sum_{i=1}^M L_i$ signals plus noise. We
now describe a protocol which allows each user to split its power and set its rates independently, and allows the
receiver to decode all virtual users one by one via a generalized successive decoding mechanism. Recall that for the
Gaussian MAC, successive decoding works as follows. Users are decoded one after another regarding all other users that
have not been decoded as interference, and the signals of decoded users are subtracted from the overall received
signal.

\vspace{0.2cm} \textit{PROTOCOL $1$}: For user~$i$, the power split and rate  allocation rule are defined as follows:
for $k = 1,...,L_i$,
\begin{align*}
p_{ik} & = \frac{N}{M}\left(1+\frac{MP}{N}\right)^{\frac{L_i-k}{L_i}}\left[\left(1+\frac{MP}{N}\right)^{\frac{1}{L_i}}-1\right],\\
r_{ik} & = \frac{1}{2} \log \left(1 + \frac{p_{ik}}{M\left(P - \sum_{j<k} p_{ij} \right) - p_{ik} + N} \right).
%& = \frac{1}{2} \log \left(1 + \frac{\frac{1}{L_i}P}{M\left(P -
%\frac{k-1}{L_i}P \right) - \frac{1}{L_i}P + N} \right)
\end{align*}
Note that the power split and rate allocation rule in \textit{PROTOCOL~$1$} are  the same as that discussed in
Section~\ref{sec:gau}.

The generalized successive decoding algorithm is given by the following  pseudo-program. Note that after a virtual user
is decoded, its signal is subtracted from the overall received signal.

\vspace{0.2cm}

\textbf{decode} virtual users $(11,21,...,M1)$ in any order or in parallel.

\textbf{set} $\mathbf{p} = \left( p_{11},p_{21},...,p_{M1}
\right)$

\textbf{while} (some virtual users are not decoded),

\hspace{1cm} Find the minimal element in $\mathbf{p}$, say the $i$th entry;

\hspace{1cm} Decode the subsequent virtual user of user~$i$;

\hspace{1cm} Update the $i$th entry of $\mathbf{p}$: $\mathbf{p}(i) = \mathbf{p}(i) + \tilde{p}$,

\hspace{1.5cm} where $ \tilde{p}$ is the power of the virtual user

\hspace{1.5cm} being decoded in the previous step;

\textbf{end} \vspace{0.2cm}

\begin{lemma}\label{le:adrs}
If all $M$ users adopt the power split and the rate allocation rule described in \textit{PROTOCOL~$1$}, then for any
$L_i \in \mathbb{Z}^+$, $i = 1,2,...,M$, the decoder can decode all virtual users one by one following the decoding
algorithm.
\end{lemma}
\vspace{0.2cm}

\textit{Proof:} By the rate allocation rule, $r_{i1} = \frac{1}{2} \log (1 + \frac{p_{i1}}{MP - p_{i1} + N} )$ for
$i=1,...,M$. Thus, each of them can tolerate the maximum amount of interference plus noise, $MP - p_{i1} + N$. It is
then easy to see that virtual users $(11,...,M1)$ can be decoded reliably in any order or in parallel. Now, we set
$\mathbf{p} = ( p_{11},...,p_{M1} )$.

In the first run of the \textbf{while} loop, if $p_{i1}$ is the minimum entry in $ \mathbf{p}$, the receiver decodes
virtual user~$i2$. By the rate allocation rule, $r_{i2} = \frac{1}{2} \log (1 + \frac{p_{i2}}{M(P - p_{i1} ) - p_{i2} +
N} )$. This implies that the maximum amount of interference plus noise that virtual user~$i2$ can tolerate is $M\left(P
- p_{i1} \right) - p_{i2} + N$. However, the real amount of interference plus noise it faces is $MP - \sum_{j=1}^M
p_{j1} - p_{i2} + N$, which is smaller than or equal to what it can tolerate because $Mp_{i1} \leq \sum_{j=1}^M
p_{j1}$. Therefore, virtual user~$i2$ can be decoded reliably at the receiver.

Suppose the decoding process succeeds in the $t$th run of the \textbf{while} loop. Now, $ \mathbf{p} \triangleq
\left(p_1^t,...,p_M^t \right)$. Let us consider the $(t+1)$th run of the \textbf{while} loop. Suppose the $i$th entry,
$p_i^t$, is the minimum entry in $\mathbf{p}$. The receiver decodes the subsequent virtual user of user~$i$, denoted by
$il$. By the rate allocation rule, $r_{il} = \frac{1}{2} \log (1 + \frac{p_{il}}{M(P - \sum_{j<l}p_{ij}) - p_{il} + N}
)$. This implies the maximum amount of interference plus noise that virtual user~$il$ can tolerate is $M(P -
\sum_{j<l}p_{ij}) - p_{il} + N$ = $M(P - p_i^t) - p_{il} + N$. However, the real amount of interference plus noise it
faces is $MP - \sum_{j=1}^M p_j^t - p_{il} +N$, which is smaller than or equal to what virtual user~$il$ can tolerate
because $Mp_i^t \leq \sum_{j=1}^M p_j^t$. Therefore, virtual user~$il$ can be decoded reliably. Hence, the lemma
follows by induction.\qed

\vspace{0.2cm}

\begin{figure}
\begin{center}
\includegraphics[width= 6cm]{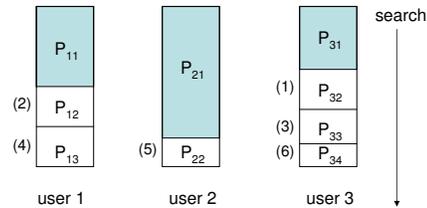}
\caption{Three-user example to illustrate the decoding algorithm. The numbers correspond to the decoding order inside the
\textbf{while} loop.}\label{fig:examgau}
\end{center}
\end{figure}

To illustrate the decoding algorithm, let us carefully examine a three-user example shown in  Fig.~\ref{fig:examgau}.
The shaded regions correspond to the virtual users $(11,21,31)$, which are decoded in any order. Suppose we decode
``$11$'' first. By the rate allocation rule, $r_{11} = \frac{1}{2} \log (1 + \frac{p_{11}}{3P - p_{11} + N} )$, which
means the maximum amount interference plus noise that virtual user~``$11$'' can tolerate is $3P -p_{11} + N$. This is
exactly the amount of interference plus noise it faces. Therefore, ``$11$'' can be decoded reliably and we can subtract
the signal of virtual user ``$11$'' from the overall received signal. Similarly, $(21,31)$ can be decoded reliably and
subtracted from the overall received signal. Now $\mathbf{p} = (p_{11},p_{21},p_{31})$. The subsequent decoding order
is illustrated by the numbers in Fig.~\ref{fig:examgau}. In the first run of the \textbf{while} loop, since $p_{31}$ is
the minimum in $\mathbf{p}$, the receiver decodes virtual user~``$32$''. By the rate allocation rule, $r_{32}
=\frac{1}{2} \log (1 + \frac{p_{32}}{3(P - p_{31}) - p_{32} + N} ) $, which implies the maximum amount interference
plus noise it can tolerate is $3(P - p_{31}) - p_{32} + N$. However, the real interference plus noise it faces is
$(P-p_{11}) + (P - p_{21}) + (P-p_{31}-p_{32}) + N$, which is smaller than what it can tolerate since
$2p_{31}<p_{11}+p_{21}$. So virtual user~``$32$'' can be decoded reliably and subtracted from the received signal. By
searching for the minimum entry in $\mathbf{p}$ in each run, we always decode a virtual user that can tolerate more
interference than what it really faces. This assures the validity of our decoding algorithm. All of the other virtual
users can be decoded in a similar fashion.

By Lemma~\ref{le:morevuser} and Lemma~\ref{le:adrs}, user~$i$ can choose any $L_i \in \mathbb{Z}^+$,  independently
from other users, and have all virtual users decoded reliably at the receiver. Therefore, user~$i$ can choose $L_i$
according to its own service requirement. For example, if user~$i$ wants to send low rate voice communication packets,
it can set $L_i = 1$, which corresponds to the basic CDMA scheme. If user~$i$ wants to send high rate stream video, it
can set $L_i$ equal to a large value in order to get higher throughput at the expense of higher coding complexity.
Thus, this variation of DRS provides an explicit way for end users to trade off throughput and coding complexity,
making differential rate requirements achievable in a distributed manner. Finally, Corollary~\ref{cor:conv} and
Theorem~\ref{thm:speed} demonstrate the asymptotic optimality of this scheme and its rate of convergence.

\subsubsection{$M$-user Discrete Memoryless MAC}

In this section, we describe a variation of the DRS scheme for the discrete memoryless MAC which supports differential
rate requirements to end users in a distributed manner. In this scheme, we adopt the uniform switch, but we do not
require
 every user to have the same number of virtual users. We split by means of $M$ independent switches. Without loss of
generality, let us consider user~$i$. If user~$i$ has $L_i$ virtual users, then switch~$i$ has $L_i$ inputs, $X_{ik}
\in {\cal X}_i$ for $k = 1,...,L_i$, and one output $X_i \in {\cal X}_i$. Switch~$i$ is controlled by a uniform random
variable $S_i \in \{1,...,L_i\}$ with $\mathsf{P} (S_i=k) = \frac{1}{L_i}$ for $k = 1,..., L_i$. The output is given
by: $ X_i = X_{ik}$, if $S_i = k$.

We now describe the protocol for the discrete memoryless MAC which allows user~$i$ to choose $L_i$  independently. We
show the asymptotic optimality of this variation of DRS under the protocol.

\textit{PROTOCOL 2}: For user~$i$ with $L_i$ virtual users, the switch~$i$ is controlled by a uniform  random variable
$S_i$ where $\mathsf{P} (S_i=k) = \frac{1}{L_i}$ for $k = 1,..., L_i$. The rate allocation rule is defined as follows:
\begin{multline*}
r_{X_{ik}} =  \frac{1}{L_i} \left[ \left(1- \frac{k-1}{L_i} \right)^{M-1} I(X_1;Y) \right.\\
 \left. + \sum_{l=1}^{M-1} {M-1 \choose l} \left(\frac{k-1}{L_i} \right)^l \cdot \right. \\
 \left. \hspace{1cm} \left(1-\frac{k-1}{L_i} \right)^{M-1-l} I(X_1;Y,X_2,..,X_{l+1}) \right].
\end{multline*}

The decoding algorithm is given by the following pseudo-program. Note that after a virtual user  is decoded, its signal
is used as side information to aid the decoding process of subsequent virtual users. \vspace{0.3cm}

\textbf{decode} virtual users $(11,21,...,M1)$ in any order or in parallel.

\textbf{set} $\mathbf{s} = (\frac{1}{L_1},\frac{1}{L_2},...,\frac{1}{L_M} )$

\textbf{while} (some virtual users are not decoded),

\hspace{1cm} Find the minimal element in $\mathbf{s}$, say the $i$th entry;

\hspace{1cm} Decode the subsequent virtual user of user $i$;

\hspace{1cm} Update the $i$th entry of $\mathbf{s}$: $\mathbf{s}(i) = \mathbf{s}(i) + \frac{1}{L_i}$;

\textbf{end}

\vspace{0.2cm}
\begin{lemma}
If all $M$ users adopt the rate allocation rule in \textit{PROTOCOL 2}, then for any $L_i \in \mathbb{Z}^+$ $\forall
i$, the decoder can decode all virtual users one by one following the decoding algorithm.
\end{lemma}
\vspace{0.1cm}

\textit{Proof}: Virtual users $(11,21,...,M1)$ can be decoded in any order if and only  if $r_{X_{i1}} \leq
I(X_{i1};Y,\mathbf{S}_1^M)$ for all~$i$. This is true because under our rate allocation rule, $r_{X_{i1}} =
\frac{1}{L_i} I(X_1;Y) = \frac{1}{L_i} I(X_i;Y) = I(X_{i1};Y,\mathbf{S}_1^M)$. We set $\mathbf{s} =
(\frac{1}{L_1},...,\frac{1}{L_M} ).$

In the first run of the \textbf{while} loop, if $\frac{1}{L_i}$ is the minimum entry of $\mathbf{s}$, the  receiver
decodes virtual user~$i2$. By the rate allocation rule
\begin{eqnarray}
r_{X_{i2}} & = & \frac{1}{L_i} \left[ \left(1- \frac{1}{L_i} \right)^{M-1} I(X_1;Y) \right. \nonumber \\
& & +  \sum_{l=1}^{M-1} {M-1 \choose l} \left(\frac{1}{L_i} \right)^l \cdot \nonumber \\
& & \left. \left(1-\frac{1}{L_i} \right)^{M-1-l} I(X_1;Y,X_2,..,X_{l+1}) \right] \label{eq:i2}
\end{eqnarray}

Virtual user~$i2$ can be decoded reliably if $r_{X_{i2}} \leq I(X_{i2};Y,\textbf{S}_1^M,X_{11},...,X_{M1} ).$ This
mutual information can be simplified in the same way as described in equations~\eqref{eq:simplify} - \eqref{eq:mlast}.
Recall the definition $-i \equiv \left\{1,...,M\right\} \backslash \left\{ i \right\}$.
\begin{eqnarray}
& & I\left(X_{i2};Y, \textbf{S}_1^M,X_{11},...,X_{M1} \right) \nonumber \\
& = & \frac{1}{L_i} I\left(X_{i2};Y,X_{11},...X_{M1} \left|
\textbf{S}_{-i}
\right. \right) \\
& = & \frac{1}{L_i} \sum_{\textbf{s}_{-i}} \mathsf{P} \left(\textbf{S}_{-i} = \textbf{s}_{-i}\right) I
\left(X_1;Y,X_{11},..., \right. \nonumber \\
& & \left. \hspace{1cm} X_{M1} \left| \textbf{S}_{-i} = \textbf{s}_{-i} \right. \right). \label{eq:star}
\end{eqnarray}
We can verify that the $I(X_1;Y)$ term in~\eqref{eq:i2} is less than or equal to the corresponding $I(X_1;Y)$ term in
Equation~\eqref{eq:star}. This follows from
\begin{eqnarray*}
& & \mathsf{P} \left(S_1 > 1,...,S_{i-1}>1,S_{i+1}>1,...S_M > 1 \right) \\ &=& \prod_{j \neq i} \left( 1 -
\frac{1}{L_j} \right) \\ &\geq& \left(1 - \frac{1}{L_i} \right)^{M-1}, \end{eqnarray*}since $\frac{1}{L_i} \leq
\frac{1}{L_j}$ for all $j \neq i$. Similarly, it can be verified that the second term in~\eqref{eq:i2} is less than or
equal to the corresponding term in Equation~\eqref{eq:star}. Therefore, virtual user~$i2$ can be decoded reliably.

Suppose the decoding process succeeds in the $t$th run of the \textbf{while} loop. Now, $\mathbf{s}\triangleq \left(
s_1^t,...,s_M^t \right)$. Let us consider the $(t+1)$th run of the \textbf{while} loop. Suppose the $i$th entry,
$s_i^t$, is the minimum entry in $\mathbf{s}$. The receiver decodes the subsequent virtual user of user~$i$, denoted by
$ij$. By the rate allocation rule
\begin{eqnarray*}
r_{X_{ij}} & = & \frac{1}{L_i} \left[ \left(1- \frac{j-1}{L_i} \right)^{M-1} I(X_1;Y) \right. \\
& &  + \sum_{l=1}^{M-1} {M-1 \choose l} \left(\frac{j-1}{L_i} \right)^l \cdot \\
& & \left. \left(1-\frac{j-1}{L_i}\right)^{M-1-l} I(X_1;Y,X_2,..,X_{l+1}) \right].
\end{eqnarray*}

Again, we can simplify the mutual information $I (X_{ij};Y, \textbf{S}_1^M,\textbf{X}_1^{j-1},...,\textbf{X}_M^{j-1})$
and show that it is great than or equal to $r_{X_{ij}}$, which implies that virtual user~$ij$ can be decoded reliably.
Hence, the lemma follows by induction. \qed

Let us illustrate the decoding algorithm by the following example. We consider a two-user discrete memoryless  MAC
where user~$1$ creates $2$ virtual users and user~$2$ creates $3$ virtual users. Random switch $1$ is controlled by
$S_1$ where $\mathsf{P}(S_1 = 1) = \mathsf{P}(S_1 = 2) = \frac{1}{2}$, and random switch~$2$ is controlled by $S_2$
where $\mathsf{P}(S_2 = k) = \frac{1}{3}$ for $k = 1,2,3$. By the rate allocation rule, the virtual users' rates can be
simplified as follows
\begin{eqnarray*}
r_{X_{11}} & = & \frac{1}{2} I(X_1;Y)\\
r_{X_{12}} & = & \frac{1}{2} \left[\frac{1}{2} I(X_1;Y) + \frac{1}{2}I(X_1;Y,X_2) \right]
\end{eqnarray*}
\begin{eqnarray*}
r_{X_{21}} & = & \frac{1}{3} I(X_1;Y)\\
r_{X_{22}} & = & \frac{1}{3}\left[\frac{2}{3} I(X_1;Y) + \frac{1}{3}I(X_1;Y,X_2)\right]\\
r_{X_{23}} & = & \frac{1}{3}\left[\frac{1}{3} I(X_1;Y) + \frac{2}{3}I(X_1;Y,X_2) \right].
\end{eqnarray*}

We first decode $(11,21)$ in any order. Suppose we decode ``$21$'' first. Virtual user ``$21$'' can be decoded reliably
if $r_{X_{21}} \leq I \left( X_{21};Y, S_1, S_2\right)$. The condition holds because
\begin{eqnarray*}
I \left(X_{21};Y, S_1, S_2\right) &=& I \left(X_{21};Y,S_1|S_2 \right) \\ & =& \frac{1}{3} I \left(X_{21};Y|S_2 =
1\right) \\ &=& \frac{1}{3} I \left(X_2;Y\right) \\ &=& \frac{1}{3} I \left(X_1;Y\right) \\ &=& r_{X_{21}}
\end{eqnarray*}
Virtual user~``$11$'' can be decoded similarly. Now, we set $\mathbf{s} = (\frac{1}{2}, \frac{1}{3} )$.

In the first run of the \textbf{while} loop, the receiver decodes ``$22$'' since $\frac{1}{3} < \frac{1}{2}$. Let  us
calculate the mutual information between $X_{22}$ and $Y$ given $S_1,S_2$ and previously decoded $X_{11},X_{21}$.

\begin{eqnarray*}
& & I \left(X_{22};Y,S_1,S_2,X_{11},X_{21}\right) \\
& = & \frac{1}{3} I \left(X_{22};Y|S_1,X_{11},X_{21},S_2 = 2\right) \\
& = & \frac{1}{3} \left(\frac{1}{2} I (X_{2};Y,X_1) + \frac{1}{2} I (X_{2};Y) \right) \\
& = & \frac{1}{3} \left(\frac{1}{2} I (X_{1};Y,X_2) + \frac{1}{2} I (X_{1};Y) \right) \\
& \geq & \frac{1}{3}\left( \frac{1}{3}I(X_1;Y,X_2) + \frac{2}{3} I(X_1;Y) \right) \\
& = & r_{X_{22}}.
\end{eqnarray*}
The third equality is due to our symmetric assumption. Therefore, virtual user~``$22$'' can be decoded. By searching
for the minimum entry in $\mathbf{s}$ in each run, we always decode a virtual user whose rate is smaller than or equal
to the corresponding mutual information. This guarantees the correctness of our decoding algorithm. Virtual users
$(12,23)$ can be decoded reliably at the receiver in a similar fashion. The asymptotic optimality of this scheme in the
discrete memoryless MAC can be demonstrated by Theorem~\ref{thm:mconv}.

\section{Concluding Remarks} \label{sec:conclusion}
In this paper, we take an information-theoretic approach to the problem of  distributed multiple-access communication.
We present a Distributed Rate Splitting scheme whereby each real user creates a number of virtual users and all virtual
users are successively decoded at the receiver. One possible advantage of Distributed Rate Splitting is that it can be
implemented with lower complexity when compared with joint coding schemes, and less coordination among users when
compared with either time-sharing or rate splitting.  For the symmetric $M$-user Gaussian MAC, each real user creates
the same number of virtual users via a power/rate splitting mechanism. The transmitted signal of a real user is the
superposition of all its virtual users' signals.  For the symmetric $M$-user discrete memoryless MAC, each real user
creates the same number of virtual users via a random switching mechanism, and the transmitted signal of a real user is
determined by a random switch. All virtual users are successively decoded at the receiver. It is shown that DRS can
achieve the maximum equal rate point for both channel models as the number of virtual users per real user tends to
infinity. Finally, we present two variations of the DRS scheme. For the case of asymmetric capacity regions, we show
that a point on the dominant face can be achieved asymptotically. For the case of an unequal number of virtual users,
we show that different user rates requirements can be accommodated independently in a distributed manner.

%\renewcommand{\theequation}{A-\arabic{equation}}
  % redefine the command that creates the equation no.
%\setcounter{equation}{0}  % reset counter
\appendices
\section{Proof of Theorem~\ref{thm:unique}} \label{sec:appendixA} % use *-form to suppress numbering

\textit{Proof:} We use induction on the number of virtual users. For $L=2$, the original problem reduces to:
$SIR_1 \left( p_1,p_2 \right) = SIR_2 \left( p_1,p_2\right)$ subject to $p_1+p_2 = P$ and $p_1,p_2 \geq 0$.
The unique solution is
\begin{eqnarray*}
(p_1^*(P),p_2^*(P))& = & \left(\frac{1}{M} \left(N+MP-\sqrt{N\left(N+MP\right)}\right), \right. \\
 & & \hspace{0.5cm} \left. \frac{1}{M} \left(-N+\sqrt{N\left(N+MP\right)} \right)\right).
\end{eqnarray*}
For $L=j-1$, suppose $(p^{*}_1(P),...,p^{*}_{j-1}(P) )$ uniquely solves $SIR_1 \left(p_1,...,p_{j-1} \right)= ... =
SIR_{j-1}\left(p_1,...,p_{j-1} \right)$ for any $P>0$, subject to $\sum^{j-1}_{k=1}p_k = P$ and $p_k \geq 0 \text{ for
} k=1,2,..,j-1$. Let us consider the $j$ virtual users case. Given any tuple $\left( p_1,...,p_j\right)$ such that
$\sum_{k=1}^j p_k = P \text{ and } p_k \geq 0 \text{ for } k=1,2,..,j$, we can fix $p_j$, so $p_1 + ... + p_{j-1} =
P-p_j$. We now solve
\begin{equation} \label{eq:hminusone}
SIR_1 \left(p_1,...,p_{j-1} \right) =... = SIR_{j-1}\left(p_1,...,p_{j-1}\right)
\end{equation}
subject to $\sum^{j-1}_{k=1}$$p_k$ = $P-p_j$, $p_k$ $\geq 0$ for $k$=$1,2,..,j.$ For fixed $p_j$, by the induction
hypothesis, we have a unique solution $( p^{*}_1$$(P-p_j)$, ...,$p^{*}_{j-1}$$(P-p_j))$ which solves
equation~(\ref{eq:hminusone}). Let $s^*\left(p_j\right) \triangleq $
$SIR_{j-1}\left(p^*_1\left(P-p_j\right),...,p^*_{j-1}\left(P-p_j\right) \right) =
\frac{p^*_{j-1}\left(P-p_j\right)}{\left(M-1\right)p^*_{j-1}\left(P-p_j\right)+N}$. We are able to verify that
$s^{*}(p_j)$ is a strictly decreasing function of $p_j$, and $SIR_j(p_j)$ = $\frac{p_j}{(M-1)p_j+N}$ is a strictly
increasing function of $p_j$. Moreover, the function values at boundary points satisfy $s^{*} \left( 0 \right)
> SIR_j \left( 0\right)$ and $s^{*} \left( P \right) < SIR_j \left( P\right)$. So there exists a unique $p_j^{*}$ such
that $SIR_j\left(p_j^{*}\right) = s^{*}\left(p_j^{*}\right)$. Hence, we conclude $\left(
p^{*}_1\left(P-p^{*}_j\right),...,p^{*}_{j-1}\left(P-p^{*}_j\right), p^{*}_j \right)$ is the unique solution to the $j$
virtual users case. The theorem follows by induction. \qed

\section*{Acknowledgment}
Edmund Yeh gratefully acknowledges useful discussions with Prof. Shlomo Shamai on this topic.

\bibliographystyle{IEEEtran}
%\bibliography{IEEEabrv,mybibfile}

\begin{biography}{Jian Cao}
Jian Cao received the B.S. degree in Electrical and Electronics Engineering with first class honor from the Hong Kong
University of Science and Technology, Hong Kong, in 2002, the M.S. degree in Electrical Engineering from Yale
University, New Haven, in 2003, and is currently working toward the Ph.D. degree in Electrical Engineering at Yale. His
research interests include information theory and queueing theory.
\end{biography}

\begin{biography}{Edmund Yeh}
Edmund Yeh received his B.S. in Electrical Engineering with Distinction from Stanford University in 1994, M.Phil in
Engineering from the University of Cambridge in 1995, and Ph.D. in Electrical Engineering and Computer Science from MIT
in 2001.  Since July 2001, he has been on the faculty at Yale University, New Haven, Connecticut, where he is currently
an Associate Professor of Electrical Engineering and Computer Science.

Dr. Yeh is a recipient of the Army Research Office (ARO) Young Investigator Program (YIP) Award (2003), the Winston
Churchill Scholarship (1994), the National Science Foundation and Office of Naval Research Fellowships (1994) for
graduate study, the Frederick E. Terman Award from Stanford University (1994) and the Barry M. Goldwater Scholarship
from the United States Congress (1993).  Dr. Yeh is a member of Phi Beta Kappa, Tau Beta Pi, and IEEE.  He has been
visiting faculty at MIT, Princeton University, University of California at Berkeley, and Swiss Federal Institute of
Technology, Lausanne.

\end{biography}
\end{document}